\documentclass[aps,twocolumn,pra]{revtex4}

\usepackage{amsmath,amssymb,bm}
\usepackage{graphicx}
\usepackage{epstopdf}
\usepackage{latexsym}
\usepackage{subfigure}
\usepackage[usenames,dvipsnames]{color}
\usepackage{hyperref}
\usepackage{natbib}

\begin{document}
\newcommand{\cs}[1]{{\color{blue}$\clubsuit$#1}}

\title{Strongly interacting fermions in an optical lattice}
\author{M. L. Wall and L. D. Carr}
\affiliation{Department of Physics, Colorado School of Mines, Golden, CO 80401, USA}

\begin{abstract}

We analyze a system of two-component fermions which interact via a Feshbach resonance in the presence of a three-dimensional lattice potential.  By expressing a two-channel model of the resonance in the basis of Bloch states appropriate for the lattice, we derive an eigenvalue equation for the two-particle bound states which is nonlinear in the energy eigenvalue.  Compact expressions for the interchannel matrix elements, numerical methods for the solution of the nonlinear eigenvalue problem, and a renormalization procedure to remove ultraviolet divergences are presented.  From the structure of the two-body solutions we identify the relevant degrees of freedom which describe the resonance behavior in the lowest Bloch band.  These degrees of freedom, which we call dressed molecules, form an effective closed channel in a many-body model of the resonance, the Fermi resonance Hamiltonian (FRH).  It is shown how the properties of the FRH can be determined numerically by solving a projected lattice two-channel model at the two-particle level.  As opposed to single-channel lattice models such as the Hubbard model, the FRH is valid for general $s$-wave scattering length and resonance width.  Hence, the FRH provides an accurate description of the BEC-BCS crossover for ultracold fermions on an optical lattice.

\end{abstract}
\maketitle

\section{Introduction}
\label{sec:FRHPRA:introduction}

In the absence of interactions, bosons cooled to zero temperature undergo a phase transition into a coherent quantum state of matter, the Bose-Einstein condensate (BEC), in which all particles reside in one single-particle state.  Fermions, on the other hand, evolve smoothly from a classical gas into a degenerate Fermi gas where the $N$ particles occupy the $N$ lowest energy single-particle states.  However, in the presence of arbitrarily weak interactions the Fermi gas has an instability towards a the Bardeen-Cooper-Schrieffer (BCS) superfluid in which fermions are bound to form bosonic Cooper pairs and these pairs then condense to form a BEC~\cite{Leggett_06}.  In this case there is a transition from the normal gas to the BCS superfluid which occurs at temperatures much lower than the characteristic temperature which marks the onset of quantum degeneracy.  As the attractive potential between fermions increases, the binding energy of a fermionic pair becomes much larger than the Fermi energy and these pairs become tightly bound bosonic molecules, a phenomenon known as the BEC-BCS crossover.  The fact that this crossover is smooth implies that we can meaningfully speak of the crossover as Bose condensation of diatomic molecules with coupling-dependent properties~\cite{Leggett_80}.  Thus, an understanding of these diatomic molecules, the two-body bound states, is key to understanding the behavior of the system throughout the crossover.

The BEC-BCS crossover has fascinated theorists for more than three decades~\cite{Eagles_69,Leggett_80,Nozieres_SchmittRink_85,Chen_Stajic_05}, but it is only within the last few years that it has been amenable to experimental study~\cite{Regal_Greiner_04b,Zwierlein_Stan_04,Barteinstein_Altmeyer_04,Kinast_Hemmer_04,Bourdel_Khaykovich_04}.  One prevalent modern context of the BEC-BCS crossover is ultracold fermionic atoms in optical traps interacting via a Feshbach resonance.  In particular, this work focuses on Feshbach interacting fermions in an optical lattice, a periodic potential made of interfering laser beams.  The lattice introduces richness into the problem which is not encountered in the continuum or harmonic traps.  For example, because of the reduced translational symmetry, fermions can pair to form molecules with a center of mass quasimomentum differing by a reciprocal lattice vector, leading to the possibility of multiple bound states for a fixed $s$-wave scattering length.  Additionally, the center of mass, relative, and internal degrees of freedom of an object in a lattice do not separate as in the continuum or a harmonic trap.  This gives rise to bound state properties which depend both on the internal state and the center of mass motion of the object.  Lastly, new many-body phases can be induced by a lattice, such as the Ising deconfinement transition between an atomic superfluid and a molecular superfluid predicted for Feshbach interacting bosons in lattices~\cite{Radzihovsky_Park_04,Radzihovsky_Weichman_08,Ejima_Bhaseen_11}.

A prevailing theme in modern condensed matter physics is that the relevant degrees of freedom in a strongly interacting many-body system are often not the microscopic degrees of freedom.  This realization is the underpinning of the Landau Fermi Liquid theory, in which quasiparticles with renormalized mass and interactions form a liquid in one-to-one correspondence with a noninteracting Fermi gas.  The renormalized parameters of such quasiparticle theories can often be determined from few-body physics by an appropriate dressing of the interaction via a Bethe-Salpeter equation or a renormalization group procedure~\cite{Shankar_94}.  In our situation the scattering of two fermions at low energies can be well described by a few scattering channels in the continuum, but in the lattice each of these channels inherits a band index.  The interchannel coupling is typically much larger than the band gap for a broad Feshbach resonance, and so all of the bands become strongly coupled\footnote{In the continuum, the only scale with which to compare the width of the resonance is the Fermi energy.  In the lattice we measure the width of the resonance by the ratio of the effective range to the lattice spacing.  For further clarification of this point, see the discussion following Eq.~\eqref{eq:infbroadNLEE}.}.  This suggests that the bare channels which are relevant in the continuum are no longer an appropriate description for the strongly correlated physics in the lattice, as we must sum over a large number of bands for an accurate solution.  Instead, we use the following facts to identify relevant dressed channels for the lattice system: (i) the interchannel coupling is large compared to the other energy scales of the problem; (ii) the interaction is short range compared to the other length scales of the problem, in particular the lattice scale; and (iii) the density per unit cell is low.  A table of scales of the lattice problem is given in Sec.~\ref{sec:EnergyScales}.

\begin{figure}[htbp]
\centerline{\includegraphics[width=0.75\columnwidth]{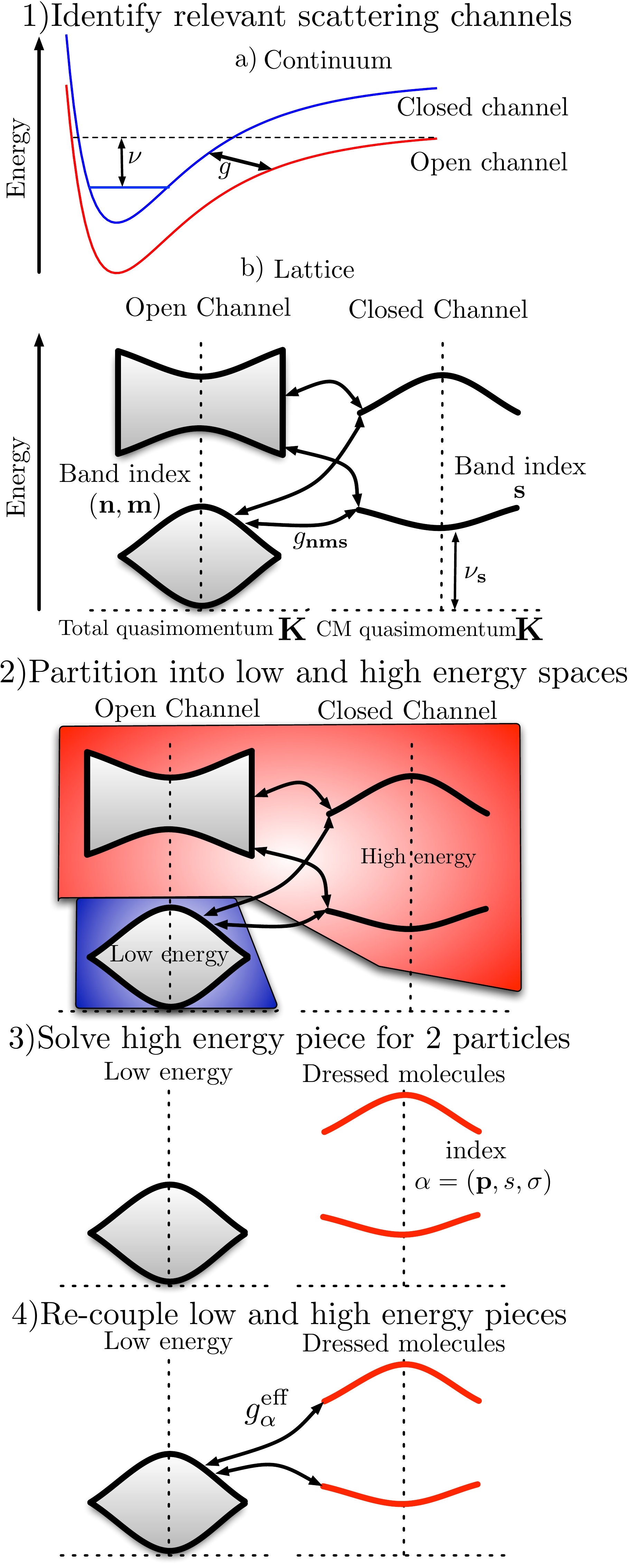}}
\caption{\label{fig:FRHPRA:schematic}(Color online)  \emph{Schematic of effective model transformation.} We model the continuum scattering of fermions using a two-channel model for the Feshbach resonance.  In the presence of the lattice, all channels inherit band indices.  The Hilbert space is partitioned into low and high energy sectors relative to the band gap, and the high energy sector is diagonalized at the two-body level.  The two-body solutions of the high-energy piece are used as an effective closed channel and re-coupled to the low energy sector at the many-body level, resulting in the Fermi resonance Hamiltonian.}
\end{figure}

Our approach to identifying the relevant degrees of freedom can be described in four steps, displayed schematically in Fig.~\ref{fig:FRHPRA:schematic}.  First, we identify the channels which are relevant to the scattering of fermions at low energies in the continuum.  In this work we consider a two-channel model for a Feshbach resonance, but the extension to multichannel models is straightforward.  While we consider only two channels, corresponding to two independent states in the continuum, these two channels become an infinite set of states in the lattice, indexed by band indices.  The open channel is spanned by the scattering states of two particles, and so is indexed by two band indices.  In one dimension (1D), the open channel states in the lattice form scattering continua which are separated by band gaps as shown schematically in Fig.~\ref{fig:FRHPRA:schematic}.  The closed channel is a bound, point-like boson, and so is indexed by a single band index, with dispersion shown schematically in Fig.~\ref{fig:FRHPRA:schematic}.  The two-channel model parameters, the coupling constant $g$ and the detuning $\nu$, which will be discussed in more detail in Sec.~\ref{sec:FRHPRA:fewbody}, also inherit band indices.  In this way all of the lattice channels become strongly coupled when $g$ is large.  Second, we partition the Hilbert space into low energy and high energy parts, where low and high energy are defined relative to the band gap.  The low energy space consists of fermions in the lowest band of the open channel, and the high energy space consists of (a) all bands of the open channel which have at least one particle not in the lowest single-particle band and (b) all bands of the closed channel.  The high energy piece is only accessed via the Feshbach coupling, which is the largest two-body energy scale.  Additionally, as the Feshbach coupling has an effective range much smaller than the lattice spacing, the high energy portion of Hilbert space is only accessed when two particles come in very close spatial proximity.  At low lattice fillings, densities of more than two particles per unit cell are rare, and so we neglect interactions involving more than two particles.  Accordingly, the third step in our effective model transformation is to numerically solve the high energy piece for two particles.  Our specific numerical method is discussed in Appendices \ref{sec:GMA} and \ref{sec:NLEEsoln}.  The dressed eigenstates of the two-particle high-energy sector of Hilbert space form a more appropriate basis for describing scattering in a lattice at low energies than the bare bands.  In the final step, we re-couple the low energy sector to the dressed high energy sector at the many-body level, resulting in a two-channel lattice model with renormalized parameters which we call the \emph{Fermi resonance Hamiltonian} (FRH)~\cite{wall_carr_12}.

There have been many other works, referenced and discussed in detail in Section \ref{sec:FRHPRA:overview}, which have attempted to identify the relevant dressed degrees of freedom for lattice fermions.  A key difference between our works and others is that the solution of the high-energy piece, step 3 in our scheme, is done using the full lattice potential and not a tight-binding or harmonic approximation.  Approximations such as these which cause artificial separation of the center of mass and relative coordinates lead incorrectly to both qualitative and quantitative differences in the effective model when the interactions are strong.

This article is organized as follows.  Section \ref{sec:FRHPRA:overview} reviews theoretical approaches to describing the BEC-BCS crossover in cold Fermi gases with a focus on the crossover in trapped geometries.  Section \ref{sec:EnergyScales} provides an overview of the physical scales of the problem.  Section \ref{sec:FRHPRA:fewbody} derives an equation for the bound states of two fermions in a 3D optical lattice interacting through a Feshbach resonance, discusses the symmetries of the solutions, gives explicit expressions for the matrix elements of the interchannel Hamiltonian in the limit of an infinite lattice, and provides details of the regularization of the theory.  Section \ref{sec:FRHPRA:manybody} uses the results of Section \ref{sec:FRHPRA:fewbody} to derive an effective many-body Hamiltonian for Feshbach interacting fermions at low energies and low density.  We conclude in Section \ref{sec:FRHPRA:conclusion}.  Some practical and numerical details concerning the two-particle solution are given in the appendices.

\section{Overview of theory of strongly interacting fermions}
\label{sec:FRHPRA:overview}

The simplest approach to the BEC-BCS crossover is to use the generalized BCS ansatz, which is valid only for weak interaction and high density, at the mean-field level for arbitrary interaction and find the chemical potential self-consistently by fixing the average number of particles~\cite{Leggett_06}.  This approach was first used by Eagles in the context of superconductivity in low carrier concentration systems~\cite{Eagles_69} and later by Leggett~\cite{Leggett_80} and Nozi\'eres and Schmitt-Rink~\cite{Nozieres_SchmittRink_85} explicitly for the BEC-BCS crossover at zero temperature and finite temperature, respectively.  The last two works demonstrated that the crossover is smooth, and that the transition temperature into the superfluid phases is smooth as a function of the coupling from the BCS to the BEC limit.  Treatment at the mean-field level is quantitatively accurate for so-called narrow resonances where the width of the resonance is much smaller than the Fermi energy~\cite{Andreev_Gurarie_04,Gurarie_Radzihovsky_07}, as this provides a dimensionless small parameter in which to perturbatively expand.  However, current experiments with ultracold atoms work in the opposite limit of resonance width much greater than the Fermi energy where the absence of any small parameter makes analysis difficult.

Further work on the BEC-BCS crossover in the cold atomic gases context noted that the atom-molecule Feshbach coupling which generates the attractive interaction in alkali gases results in a condensate of molecules which is mutually coherent with the Cooper paired fermions~\cite{Timmermans_Furuya_01}.  This novel many-body state, called a resonance superfluid, has a transition temperature near the divergence of the scattering length which is larger than that of the pure BEC or BCS limits.  Models for resonance superfluids~\cite{Holland_Kokkelmans_01,Kokkelmans_Milstein_02,Timmermans_Furuya_01,Milstein_Kokkelmans_02,Ohashi_Griffin_02,Stajic_Milstein_04} include a pairing term of the schematic form
\begin{align}
&\hat{H}_{\mathrm{atom-molecule}}=\int d\mathbf{r}d\mathbf{r}'g\left(\mathbf{r}-\mathbf{r}'\right)\\
\nonumber &\times\left[{\hat{\Psi}^{\left(b\right)}}\,^{\dagger}\left(\frac{\mathbf{r}+\mathbf{r}'}{2}\right)\hat{\Psi}^{\left(f\right)}_{\uparrow}\left(\mathbf{r}\right)\hat{\Psi}^{\left(f\right)}_{\downarrow}\left(\mathbf{r}'\right)+\mathrm{h.c.}\right]\, ,
\end{align}
where $\hat{\Psi}_{\sigma}^{\left(f\right)}\left(\mathbf{r}\right)$ is a field operator for a fermion with spin $\sigma\in\left\{\uparrow,\downarrow\right\}$ and $\hat{\Psi}^{\left(b\right)}\left(\mathbf{r}\right)$ is a field operator for a bosonic molecule.  Such a term converts a pair of fermionic atoms in opposite spin states into a molecule at the center of mass, and vice versa.  In addition, to describe a Feshbach resonance one includes an energetic detuning $h\nu$ between the open channel fermions and the bosonic closed channel molecules.  To our knowledge the first time such terms were used in a cold atom context was in Ref.~\cite{Timmermans_Tommasini_99}, although such terms had been used phenomenologically in the study of high-$T_c$ superconductivity for many years prior~\cite{Ranninger_Robaszkiewicz_85,Friedberg_Lee_89,Ranninger_Robin_95}. The analogous high-$T_c$ model, known as the Cooperon or boson-fermion model, is still a subject of current research~\cite{Micnas_07,Yang_Kozik_11}.

The approximation that the pairing of two fermions to form a boson occurs only at the center of mass introduces an ultraviolet divergence in the theory.  The proper renormalization of this divergence for the two-channel resonance Hamiltonian was first carried out in the absence of a lattice by Kokkelmans {\it et al.}~\cite{Kokkelmans_Milstein_02}.  Here, two-channel means that only a single bosonic field couples to the fermions.  Kokkelmans {\it et al.} presented general relations between the bare and renormalized properties of the two-channel resonance model in terms of a momentum cutoff $K^{\star}$.  Of particular interest is the renormalization of the detuning, in which the physical detuning ${\nu}$ is related to the divergent bare detuning $\nu_0$ as
\begin{align}
\label{eq:renormalization} h{\nu}&=h\nu_0-\sum_{\mathbf{k}}\frac{g_{\mathbf{k}}^2}{2\epsilon_{\mathbf{k}}}\, ,
\end{align}
with $g_{\mathbf{k}}$ the matrix element coupling two fermions with relative momentum $\mathbf{k}$ to a boson and $\epsilon_{\mathbf{k}}=\hbar^2k^2/2m$ is the free fermion dispersion.  For the contact interaction, $g_{\mathbf{k}}=g$, and the renormalization Eq.~\eqref{eq:renormalization} becomes~\cite{Kokkelmans_Milstein_02}
\begin{align}
\label{eq:freerenorm} h{\nu}&=h\nu_0-\frac{mK^{\star}g^2}{2\pi^2\hbar^2}\, .
\end{align}
This renormalized detuning defines the zero-energy limit of the $T$-matrix~\cite{HollandProceedings,HMV}, 
\begin{align}
\label{eq:zeroETlimit}T=\frac{g^2}{h{\nu}}\, .
\end{align}
Equation~\eqref{eq:zeroETlimit} provides the correct behavior of the $s$-wave scattering length near a Feshbach resonance for any detuning, indicating that the renormalized two-channel model is an appropriate description of the resonance.  In Sec.~\ref{sec:properreg}, we discuss how to properly renormalize the lattice theory.

\begin{figure}[thbp]
\centerline{\includegraphics[width=0.9\columnwidth]{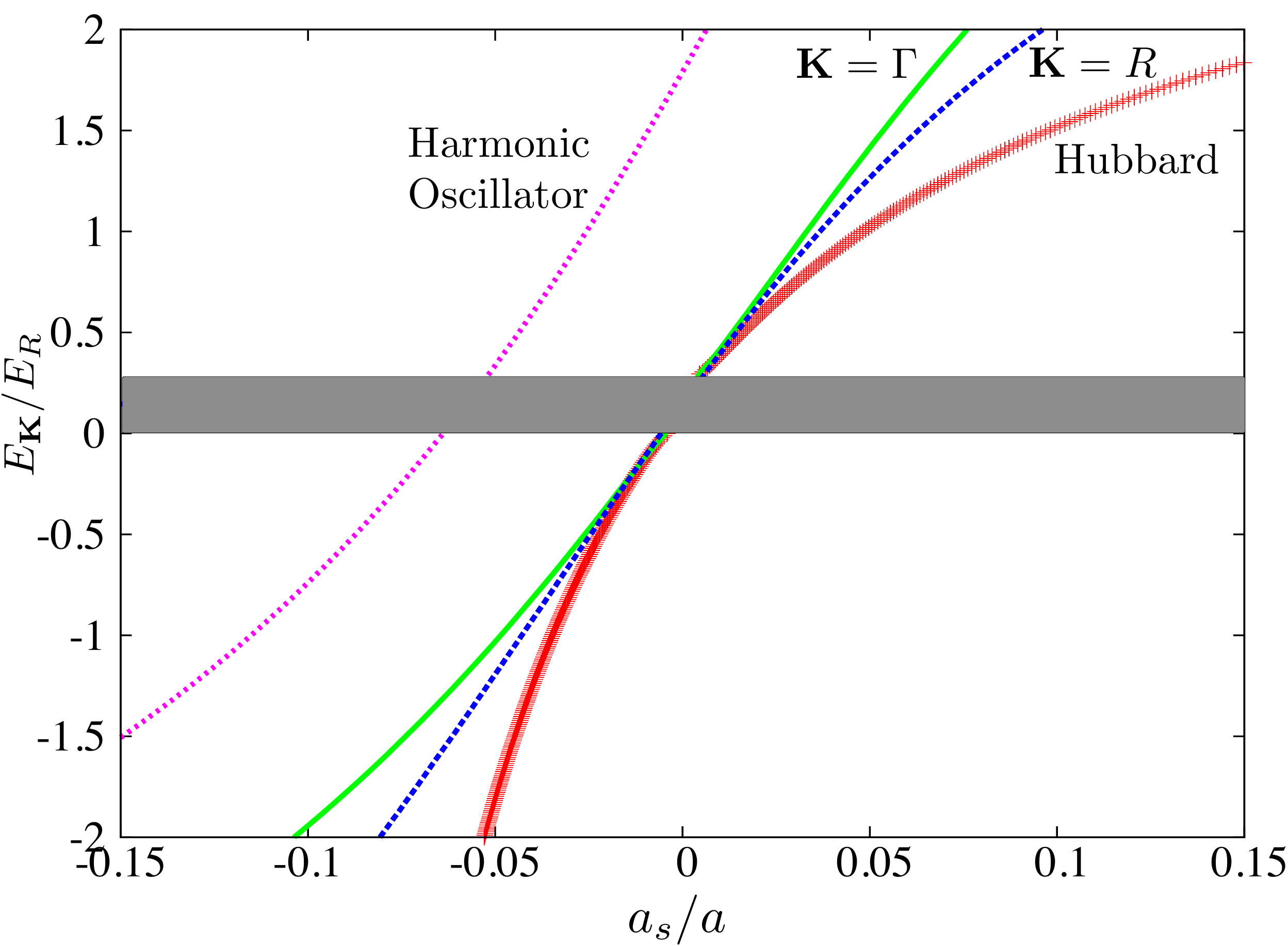}}
\caption{\label{fig:HubbComp} (Color online) \emph{Comparing approximate two-body binding energies with the numerical solution.}  The predicted two-body bound state energies of the self-consistent single-band Hubbard model\cite{Buechler_10,Buechler_12} (red points) and the harmonic oscillator approximation~\cite{Diener_Ho_06} (magenta dotted line) are compared to the numerical bound state energy of the lattice two-channel model for total quasimomentum at the center ($\Gamma$ point, solid green line) and edge ($R$ point, blue dashed line) of the Brillouin zone as a function of the $s$-wave scattering length $a_s$ for a lattice height $V=12E_R$, see Eq.~\eqref{eq:Lattdef}.  The grey stripe represents the scattering continuum of the lowest open channel scattering band.  We see significant deviations of the Hubbard model from the true solution when $a_s$ is only a few percent of the lattice spacing $a$.  No appreciable dependence of the bound state energy on quasimomentum can be seen for the Hubbard model or the harmonic oscillator, in contrast with the exact prediction.  These errors propagate to the many-body models of Refs.~\cite{Dickerscheid_Khawaja_05,Duan_05,Duan_08}, which are based on harmonic approximations of a single lattice site.}
\end{figure}

It should be stressed that the low-energy limit of the two-channel resonance theory, Eq.~\eqref{eq:zeroETlimit}, involves only renormalized quantities.  An alternative and more conventional  formulation for finding the effective interaction between fermions near a resonance is to integrate out the bosonic field using a Hubbard-Stratonovich transformation.  However, the theory resulting from a Hubbard-Stratonovich transformation is expressed solely in terms of bare rather than renormalized quantities, and so the two approaches are inequivalent.   Another means of integrating out the closed channel is replace the two-channel model with an effective single-channel model.  In the single-channel model, a contact pseudopotential interaction between fermions is introduced, and is chosen to reproduce the proper scattering length.  Such a pseudopotential causes difficulties near resonance where the scattering length diverges, as the $T$-matrix also diverges.  In the two-channel model, the effective potential between fermions is the separable potential
\begin{align}
U_{\mathbf{k},\mathbf{k}'}&=\mathcal{P}\frac{g_{\mathbf{k}}g_{\mathbf{k}'}}{2\epsilon_{\mathbf{k}}-E}\, ,
\end{align}
where $E$ is a solution of
\begin{align}
E&=h\nu_0-\mathcal{P}\sum_{\mathbf{k}}\frac{g_{\mathbf{k}}^2}{2\epsilon_{\mathbf{k}}-E}\, ,
\end{align}
and $\mathcal{P}$ denotes the Cauchy principal value~\cite{HollandProceedings,HMV}.  Hence, the two-channel model has residual momentum dependence in the effective potential even in the limit of a contact coupling $g_{\mathbf{k}}\to g$.  This momentum dependence makes the theory well-defined even at ${\nu}=0$.  The residual momentum dependence in the effective fermion interaction predicted by two-channel models is an important difference with single-channel models, see Fig.~\ref{fig:HubbComp}.

In the presence of a lattice the natural choice of single-particle states are Bloch functions, and so an expansion of the field operators in terms of Bloch functions 
\begin{align}
\hat{\Psi}_{\sigma}^{\left(f\right)}\left(\mathbf{r}\right)&=\sum_{\mathbf{n}}\sum_{\mathbf{q}\in\mathrm{BZ}}\hat{a}_{\mathbf{nq}\sigma}\phi_{\mathbf{nq}\sigma}^{\left(f\right)}\left(\mathbf{r}\right)\, ,\\
\hat{\Psi}^{\left(b\right)}\left(\mathbf{r}\right)&=\sum_{\mathbf{n}}\sum_{\mathbf{q}\in\mathrm{BZ}}\hat{b}_{\mathbf{n}\mathbf{q}}\phi_{\mathbf{n}\mathbf{q}}^{\left(b\right)}\left(\mathbf{r}\right)\, ,
\end{align}
is appropriate.  Here, $\phi_{\mathbf{nq}\sigma}^{\left(f\right)}\left(\mathbf{r}\right)$ is a Bloch function diagonalizing the fermionic single-particle Hamiltonian with $\mathbf{n}$ a band index and $\mathbf{q}$ a quasimomentum index, $\hat{a}_{\mathbf{nq}\sigma}$ creates a particle in state $\phi_{\mathbf{nq}\sigma}^{\left(f\right)}\left(\mathbf{r}\right)$, $\phi_{\mathbf{nq}}^{\left(b\right)}\left(\mathbf{r}\right)$ is a Bloch function diagonalizing the single-boson Hamiltonian, $\hat{b}_{\mathbf{n}\mathbf{q}}$ creates a particle in state $\phi_{\mathbf{n}\mathbf{q}}^{\left(b\right)}\left(\mathbf{r}\right)$, and BZ denotes the first Brillouin zone.  Using these expansions in the Hamiltonians describing resonance superfluidity leads to Fermi-Bose Hubbard Hamiltonians (FBHHs) of the generic form
\begin{align}
\nonumber \hat{H}=&\sum_{\mathbf{n}}\sum_{\mathbf{q}\in\mathrm{BZ}}E_{\mathbf{n}\mathbf{q}}^{\left(f\right)}\hat{n}_{\mathbf{n}\mathbf{q}}^{\left(f\right)}+\sum_{\mathbf{n}}\sum_{\mathbf{q}\in\mathrm{BZ}}\left(E_{\mathbf{n}\mathbf{q}}^{\left(b\right)}-h\nu\right)\hat{n}_{\mathbf{n}\mathbf{q}}^{\left(b\right)}\\
\nonumber&+\sum_{\mathbf{m},\mathbf{n},\mathbf{s}}\sum_{\mathbf{q},\mathbf{K}}\left[g_{\mathbf{sK}}^{\mathbf{n}\mathbf{m}}\left(\mathbf{q}\right)\hat{b}_{\mathbf{s}\mathbf{K}}^{\dagger}\hat{a}_{\mathbf{n}\mathbf{q}\uparrow}\hat{a}_{\mathbf{m},\mathbf{K}-\mathbf{q},\downarrow}+\mathrm{h.c.}\right]\\
\label{eq:lattresonance}&+\mbox{background terms}
\end{align}
where we have made the definitions $\hat{n}_{\mathbf{n}\mathbf{q}}^{\left(f\right)}\equiv\sum_{\sigma\in\left\{\uparrow,\downarrow\right\}}\hat{a}_{\mathbf{n}\mathbf{q}\sigma}^{\dagger}\hat{a}_{\mathbf{n}\mathbf{q}\sigma}$ and $\hat{n}_{\mathbf{n}\mathbf{q}}^{\left(b\right)}\equiv\hat{b}_{\mathbf{n}\mathbf{q}}^{\dagger}\hat{b}_{\mathbf{n}\mathbf{q}}$, $\nu$ is the detuning between the open and closed channels, $E_{\mathbf{n}\mathbf{q}}^{\left(f\right)}$ is the single-particle energy of a fermion in Bloch state $\phi_{\mathbf{n}\mathbf{q}\sigma}^{\left(f\right)}\left(\mathbf{r}\right)$, $E_{\mathbf{n}\mathbf{q}}^{\left(b\right)}$ is the single-particle energy of a boson in state $\phi_{\mathbf{n}\mathbf{q}}^{\left(b\right)}\left(\mathbf{r}\right)$, and ``background terms" denotes spatially local interaction terms such as appear in the Hubbard model~\cite{Esslinger_10} arising from the background scattering length.

Lattice FBHHs such as Eq.~\eqref{eq:lattresonance} generically require a large number of bands for an accurate solution.  Because current methods used to solve models such as the FBHH scale poorly with the number of bands, many authors have attempted to derive effective models from a FBHH which are valid in some limiting case.  For example, Carr and Holland~\cite{Carr_Holland_05} take a lowest band approximation of a FBHH, and show that in the limit of perturbative tunneling it becomes an XXZ model describing tunneling of paired fermions.  Zhou~\cite{Zhou_05} begins with a similar lowest-band model, and demonstrates that Mott insulating states are unstable with respect to fermion-boson conversion at the mean-field level.  Other authors do not take a single-band approximation, but rather group the excited fermion bands as well as the closed channel molecules into a single dressed molecule field whose properties are fixed by two-body physics.  We shall refer to such Hamiltonians as dressed effective Hubbard Hamiltonians (DEHHs).  We will review several such approaches below to contrast with the methodology taken in the remainder of this work.

Dickersheid \emph{et.~al.}~\cite{Dickerscheid_Khawaja_05} consider a DEHH, and determine the properties of the dressed molecule by considering deep lattices and replacing the lattice with a single harmonic well.  The harmonic trap approximation is the source of two systematic errors.  First it artificially leads to separability of the center of mass motion from both the relative motion and internal structure.  Second, it underestimates the extent of Wannier functions, often by an order of magnitude.  We will show that qualitative properties of the tunneling, as well as its general order of magnitude, cannot be accounted for using this approach.

The exact solution for two particles interacting via a Feshbach resonance in an anisotropic harmonic trap was determined by Diener and Ho~\cite{Diener_Ho_06}.  In their calculations they stressed the importance of intra- as well as inter-band coupling terms, and properly renormalized the theory to remove divergences from using a point-like boson.  Our results for the bound state energies reduce to theirs in the appropriate limit of deep lattices, see Fig.~\ref{fig:HubbComp}.  However, their approach is still incorrect for determining the Hubbard parameters in a DEHH, as the harmonic oscillator approximation suffers from the same systematic errors as Dickershied \textit{et al.}.  For example, the error in the tunneling is 70\% for moderate lattice depths in use in experiments.

Duan has derived DEHHs using both a projection operator formalism~\cite{Duan_05} and general symmetry considerations~\cite{Duan_08}.  In his approach, the dressed molecule is determined by the solution of a two-atom, multi-band Schr\"odinger equation governing the atom-molecule conversion process on a single site.  He then argues that when one of the dressed molecular energies is close to the scattering threshold of two particles in band $n$ the only allowed on-site configurations are empty sites, sites with a single fermion in band $n$, and sites with a single dressed molecule.  When this is the case, the Hamiltonian can be projected into the reduced Hilbert space with only these on-site configurations.   Because on-site pairing has already been taken into account in the definition of the dressed molecules, pairing in the effective many-body Hamiltonian occurs only between dressed molecules and fermion pairs which do not reside all on the same site.  His work does not give a prescription for solving or renormalizing the on-site problem, but references Diener and Ho's work with the harmonic oscillator approximation, which, as we stated, suffers from systematic errors in the calculation of the DEHH parameters.  More recent work by Kestner and Duan~\cite{Kestner_Duan_10} uses the numerical solution from a double-well potential to avoid some of the shortcomings of the harmonic oscillator approximation.  While this work captures some of the physics of the lattice at the nearest-neighbor level, it does not capture the full quasimomentum dependence of the lattice Hamiltonian.  Other approaches~\cite{Grishkevich_Sala_11,Mentink_Kokkelmans_09} have also taken partial account of anharmonic terms by series or perturbation expansions.

B\"{u}chler was the first to give the exact solution for two fermions interacting through a zero-range Feshbach resonance in an optical lattice, properly accounting for the effects of higher bands and renormalization~\cite{Buechler_10}.  He then showed that when the interaction term $U$ of the single-band Hubbard model was determined from the scattering properties of this exact two-body solution self-consistently the Hubbard model still failed to reproduce the correct physics even for moderate $s$-wave scattering length~\cite{Buechler_10,Buechler_12}.  Specifically, the prediction of the two-body bound state energy from the self-consistent single-band Hubbard model deviates significantly from the exact result for $a_s/a\gtrsim 0.02$ on the BEC side and for $|a_s|/a\gtrsim 0.06$ on the BCS side, where $a_s$ is the $s$-wave scattering length and $a$ the lattice constant, see Fig.~\ref{fig:HubbComp}.  This has motivated our approach of using a two-channel lattice model instead of a single-channel lattice model such as the Hubbard model.  As in our discussion of single-channel vs.~two-channel models in free-space, the use of a two-channel model in the lattice allows us to capture the complete quasimomentum dependence of the scattering amplitude rather than only its low energy, low-momentum behavior.  B\"{u}chler's discussion of the two-body solution focused on states with zero total quasimomentum, although the theory encompasses states with arbitrary total quasimomentum.  In this work we show how extensions of his work may be used to derive a DEHH.

Very recent work by von Stecher \emph{et.~al.}~\cite{von_Stecher_Gurarie_11} focuses on a DEHH near a lattice resonance.  Instead of solving an on-site problem they (i) project the two-body Hamiltonian outside of the scattering continuum of two fermions in bands $n$ and $m$ which gives rise to the resonance, (ii) solve this projected Hamiltonian exactly, and then (iii) use the eigenstates of this projected Hamiltonian as a dressed closed channel.  This approach is very similar in spirit to ours.  However, von Stecher \emph{et al.} assume low dimensionality (quasi-1D) from the outset.  Their approach breaks down when the energy associated with the Feshbach coupling becomes larger than the energy associated with the transverse confinement.  In contrast, our model applies to arbitrary dimensionality.  Our model treats the population in transverse excited states as being fixed by the two-body solution and thus part of the dressed molecule.  Thus, any imposed condition of reduced dimensionality, either quasi-1D or quasi-2D, can be controlled by the transverse tunneling and coupling rates of the dressed molecules.

Finally, we note one other recent work by Titvinidze and coworkers~\cite{Titvinidze_Snoek_11} which accounts for the effects of higher bands in a mean-field approximation assuming that the population of excited bands is small, and gives some predictions of the resulting model using dynamical mean-field theory.

\section{Energy scales of the lattice Feshbach problem}
\label{sec:EnergyScales}

To date, the most successful experimental realizations of two-component fermionic atoms exhibiting Feshbach resonant phenomena are $^{40}$K~\cite{greiner2003,Regal_Greiner_04,Regal_Greiner_04b,stoferle2006} and $^6$Li~\cite{strecker2003,Cubizolles_Bourdel_03,Jochim_Bartenstein_03,jochim2003b,zwierlein2003}.  Typical operating temperatures are $T\sim 100$nK in order to achieve quantum degeneracy at stable densities.  In these systems, the two internal components of the atoms correspond to two sublevels $M_F$ in a manifold of fixed total atomic spin $F$.  Feshbach resonances~\cite{Chin_Grimm_10} in these systems are due to hyperfine couplings between scattering states and a bound molecular state of the interatomic potential, and are tunable by magnetically shifting the position of the scattering threshold with respect to the bound state energy.  Such resonances can be from milligauss to hundreds of gauss wide, depending on the species and state; in practice a well-controlled uniform magnetic field is sufficient to achieve a Feshbach resonance in experiments on many atomic species.  The magnitude of the effective range $R^{\star}$ of typical Feshbach resonances in these systems, which is related to the $r_B$ which appears in the low-energy scattering amplitude Eq.~\eqref{eq:tcscattamp} in Sec.~\ref{sec:FRHPRA:fewbody} below as $R^{\star}=-2r_b$, is typically of order 20 angstroms or less~\cite{Petrov_Salomon_05}.

The lattice introduces a new length scale to the problem, the lattice constant $a$.  For optical lattices in the simple cubic arrangements that we consider in this paper, $a=\lambda/2$, where $\lambda$ is the wavelength of the retro-reflected laser light forming the lattice.  The energy scale associated with the lattice spacing is the recoil energy, $E_R=\hbar^2\pi^2/2ma^2$, which is of order $25$kHz $\approx 1\mu$K for $^{40}$K and $170$kHz $\approx 8 \mu $K for $^6$Li.  The tunneling energy of a single particle in the lowest Bloch band along a principal axis of the lattice may be expressed as
\begin{align}
\label{eq:tunnfit}\frac{t}{E_R}=1.363\left(\frac{V}{E_R}\right)^{1.057}e^{-2.117\sqrt{V/E_R}}\, ,
\end{align}
where $V$ is the lattice strength defined in Eq.~\eqref{eq:Lattdef}.  The coefficients in Eq.~\eqref{eq:tunnfit} represent the best fit to numerical data valid to $\le 1\%$ for $V>2E_R$.  The tunneling gives a measure of the width of the lowest Bloch band.  The gap between the lowest two bands in a simple cubic optical lattice is nonzero for $V\gtrsim2.4E_R$, and monotonically increases with $V/E_R$.  Near $V=12E_R$, where the numerical examples in this work are performed, the band gap is approximately $5E_R$.  The fact that the effective range is much smaller than the lattice spacing, $r_B/a\ll 1$ implies that the coupling parameter $g$ which describes the interaction between channels in the two-channel model is large in the natural units of the lattice, see Eqs.~(\ref{eq:natunits1}-\ref{eq:natunits2}) in Sec.~\ref{sec:FRHPRA:fewbody} below.  The scales of the problem are summarized in Table \ref{table:scales}.

\begin{table}[h]
\begin{center}
\begin{tabular}{|c|c|}
\hline Scale&Typical values\\
\hline\hline Lattice spacing $a$ (nm)&550\\
\hline Temperature $T$ (nK)& $\sim$100\\
\hline Recoil energy $\hbar^2\pi^2/2ma^2$ (kHz)& 25 ($^{40}$K), 170 ($^{6}$Li)\\
\hline Lowest band tunneling $t$ ($E_R$)&$\sim$0.01\\
\hline Band gap ($E_R$)&$\sim$ 5\\
\hline Effective range $r_B$ (nm)& $\lesssim$1 \\
\hline Interchannel coupling $g/E_Ra^{3/2}$&$\gtrsim $16 \\
\hline 
\end{tabular}
\caption{\label{table:scales} \emph{Table of scales of the lattice Feshbach problem.}  The values given for the tunneling and band gap refer to $V=12E_R$, see also Eq.~\eqref{eq:tunnfit}.}
\end{center}
\end{table}

\section{Two-particle solution}
\label{sec:FRHPRA:fewbody}

\subsection{Derivation of the Nonlinear eigenvalue equation}
In this section we derive an equation for the bound states of two fermions interacting through a Feshbach resonance in the presence of a 3D optical lattice~\cite{Buechler_10,wall_carr_12}.  We begin by modeling the Feshbach resonance using a two channel model in the continuum.  In this model, we partition our Hilbert space into open and energetically closed channels, with the asymptotic limit of the open channel potential corresponding to two free atoms.  Formally, this is achieved using a projector $\hat{P}$ into the open channel and its complement $\hat{Q}=\hat{1}-\hat{P}$:
\begin{align}
\label{eq:twochannel1}\left(E-\hat{H}_{PP}\right)\hat{P}\Psi&=\hat{H}_{PQ}\hat{Q}\Psi\, ,\\
\label{eq:twochannel2}\left(E-\hat{H}_{QQ}\right)\hat{Q}\Psi&=\hat{H}_{QP}\hat{P}\Psi\, ,
\end{align}
where $\hat{H}$ is the complete two-particle Hamiltonian, $\Psi$ is the two-particle wavefunction, and, e.g., $\hat{H}_{PP}=\hat{P}\hat{H}\hat{P}$ is the piece of $\hat{H}$ taking the open channel Hilbert space to the open channel Hilbert space~\cite{Feshbach_62}.  The closed channel potential $\hat{H}_{QQ}$ is assumed to support a bound molecular state near the scattering threshold of the open channel potential.  We take the open channel to be spanned by states of two fermions in different internal states with equal mass $m_1=m_2=m$ and the closed channel to be spanned by molecular states with twice the fermionic mass and twice the fermionic polarizability~\cite{Derr_Volz_06}.  We denote the two internal states of the open channel by the pseudospin notation $\left\{\uparrow,\downarrow\right\}$.  The two-channel model is defined by an interchannel coupling $g$ which pairs the open channel fermions to a closed channel molecule at the center of mass and a detuning $h\nu$ between the bound state of the closed channel and the scattering threshold of the open channel.  The two-channel scattering amplitude is~\cite{Petrov_Salomon_05}
\begin{align}
\label{eq:tcscattamp}f\left(\mathbf{k}\right)&=-\frac{\mu}{2\pi\hbar^2}\frac{g^2}{\epsilon_{\mathbf{k}}-h\nu+\frac{\mu g^2}{2\pi\hbar^2}ik}\, ,
\end{align}
where $\mu$ is the reduced mass, $\mathbf{k}$ the incident momentum, and $\epsilon_{\mathbf{k}}$ the free particle energy.  This scattering amplitude may be written in the asymptotic form of scattering of particles with momentum small compared to the inverse effective range~\cite{Landau_Lifshitz_77} by identifying the $s$-wave scattering length $a_s=-2\mu g^2/8\pi^2\hbar^3\nu$ and the effective range $r_B=\pi\hbar^4/\mu^2g^2$.  Both $a_s$ and $r_B$ can be measured experimentally, for example by radio-frequency spectroscopy of Feshbach molecules~\cite{Claussen_Kokkelmans_03}.

Denoting the wave function of the two fermions in the open channel as $\psi\left(\mathbf{x},\mathbf{y}\right)$ and the wave function of the closed channel molecules as $\phi\left(\mathbf{z}\right)$, the two-channel model, Eqs.~(\ref{eq:twochannel1}-\ref{eq:twochannel2}), in position representation becomes
\begin{align}
\nonumber[E-\hat{H}_0\left(\mathbf{x}\right)-&\hat{H}_0\left(\mathbf{y}\right)]\psi\left(\mathbf{x},\mathbf{y}\right)=\\
\label{eq:SE1} &g\int d\mathbf{z}\,\alpha\left(\mathbf{r}\right)\phi\left(\mathbf{z}\right)\delta\left(\mathbf{z}-\mathbf{R}\right)\, ,\\
\nonumber [E-h\nu_0-&\hat{H}_0^M\left(\mathbf{z}\right)]\phi\left(\mathbf{z}\right)=\\
\label{eq:SE2} &g\int d\mathbf{x}d\mathbf{y}\,\alpha\left(\mathbf{r}\right)\psi\left(\mathbf{x},\mathbf{y}\right)\delta\left(\mathbf{z}-\mathbf{R}\right)\, .
\end{align}
In this expression $\hat{H}_0\left(\mathbf{x}\right)=-\frac{\hbar^2}{2m}\nabla_{\mathbf{x}}^2+V_{\mathrm{latt}}\left(\mathbf{x}\right)$ is the single particle Hamiltonian for the open channel, $\hat{H}_0^M\left(\mathbf{z}\right)=-\frac{\hbar^2}{4m}\nabla_{\mathbf{z}}^2+2V_{\mathrm{latt}}\left(\mathbf{z}\right)$ is the single particle Hamiltonian for the closed channel, $\alpha\left(\mathbf{r}\right)$ is a regularization of the Feshbach coupling with cutoff $\Lambda$, and we have introduced the center of mass and relative coordinates $\mathbf{R}\equiv\frac{\mathbf{x}+\mathbf{y}}{2}$ and $\mathbf{r}\equiv \mathbf{x}-\mathbf{y}$.  The subscript $0$ in $\nu_0$ denotes that this is a bare detuning entering the microscopic theory which is related to the physically observable detuning $\nu$ in the limit as the cutoff $\Lambda\to \infty$ where $\alpha\left(\mathbf{r}\right)\to\delta\left(\mathbf{r}\right)$.  We will discuss our particular choice of regularization in Sec.~\ref{sec:MatrixElem}.

The eigenfunctions of the single particle Hamiltonians are Bloch functions
\begin{align}
\label{eq:Bloch1}\hat{H}_0\left(\mathbf{x}\right)\phi^{\left(f\right)}_{\mathbf{n}\mathbf{q}}\left(\mathbf{x}\right)&=E_{\mathbf{n}\mathbf{q}}^{\left(f\right)}\phi^{\left(f\right)}_{\mathbf{n}\mathbf{q}}\left(\mathbf{x}\right)\, ,\\
\label{eq:Bloch2}\hat{H}_0^M\left(\mathbf{z}\right)\phi_{\mathbf{s}\mathbf{K}}^{\left(b\right)}\left(\mathbf{z}\right)&=E_{\mathbf{s}\mathbf{K}}^{\left(b\right)}\phi_{\mathbf{s}\mathbf{K}}^{\left(b\right)}\left(\mathbf{z}\right)\, ,
\end{align}
which are described by a band index $\mathbf{n}=\left(n_x,n_y,n_z\right)$ and a quasimomentum $\mathbf{q}=\left(q_x,q_y,q_z\right)$ in the first Brillouin zone (BZ).  We establish the conventions that $\mathbf{n}$ and $\mathbf{m}$ are band indices for the open channel and their sums run from 1 to $\infty$; $\mathbf{s}$ and $\mathbf{t}$ are band indices for the closed channel and their summations run from 1 to $\infty$; $\mathbf{q}$ is a single-particle quasimomentum; $\mathbf{K}$ is the total quasimomentum; and sums over quasimomenta always run over the allowed values in the BZ.  Because the interaction is invariant under translation of the coordinate system by a Bravais lattice vector, the total quasimomentum $\mathbf{K}=\mathbf{q}_1+\mathbf{q}_2$ is conserved and we may diagonalize the Schr\"odinger equation in subspaces of fixed $\mathbf{K}$.  The open channel solution with total quasimomentum $\mathbf{K}$ may be parameterized as
\begin{align}
\label{eq:openchannel}\psi_{\mathbf{K}}\left(\mathbf{x},\mathbf{y}\right)&=\frac{1}{\sqrt{N^3}}\sum_{\mathbf{n}\mathbf{m}}\sum_{\mathbf{q}}\varphi_{\mathbf{nm}}^{\mathbf{K}}\left(\mathbf{q}\right)\phi_{\mathbf{n}\mathbf{q}}^{\left(f\right)}\left(\mathbf{x}\right)\phi_{\mathbf{m},\mathbf{K}-\mathbf{q}}^{\left(f\right)}\left(\mathbf{y}\right)\, ,
\end{align}
where $N^3$ is the total number of unit cells in a 3D lattice with periodic boundary conditions.  This wave function generalizes the idea of single-particle Bloch states to multi-particle systems, as translating the system through a Bravais lattice vector $\mathbf{R}$ multiplies the wave function by a unimodular factor $e^{-i\mathbf{K}\cdot\mathbf{R}}$.  Using this form for the open channel wave function, parameterizing the closed channel wave function as a sum over Bloch states in different bands as
\begin{align}
\label{eq:molexpansion}\Phi_{\mathbf{K}}\left(\mathbf{z}\right)&=\sum_{\mathbf{s}}\Upsilon_{\mathbf{s}}^{\mathbf{K}}\phi_{\mathbf{s}\mathbf{K}}^{\left(b\right)}\left(\mathbf{z}\right)\, ,
\end{align}
and inserting into Eqs.~(\ref{eq:SE1}-\ref{eq:SE2}), we find
\begin{align}
\label{eq:formereqn}[E_{\mathbf{K}}-E_{\mathbf{nm}}^{\mathbf{K}}\left(\mathbf{q}\right)&]\varphi_{\mathbf{nm}}^{\mathbf{K}}\left(\mathbf{q}\right)=\\
\nonumber &\frac{g}{\sqrt{v}}\sum_{\mathbf{s}}h_{\mathbf{s}\mathbf{K}}^{\mathbf{nm}}\left(\mathbf{q}\right)\Upsilon_{\mathbf{s}}^{\mathbf{K}}\, ,\\
\label{eq:lattereqn}[E_{\mathbf{K}}-h\nu_0-&E^{\left(b\right)}_{\mathbf{s}\mathbf{K}}]\Upsilon_{\mathbf{s}}^{\mathbf{K}}=\\
\nonumber&\frac{1}{N^3}\frac{g}{\sqrt{v}}\sum_{\mathbf{n}\mathbf{m};\mathbf{q}}{h_{\mathbf{s}\mathbf{K}}^{\mathbf{nm}}}^{\star}\left(\mathbf{q}\right)\varphi_{\mathbf{nm}}^{\mathbf{K}}\left(\mathbf{q}\right)\, .
\end{align}
Here, we have made the definitions of the open channel energy $E_{\mathbf{nm}}^{\mathbf{K}}\left(\mathbf{q}\right)=E_{\mathbf {n}\mathbf{q}}^{\left(f\right)}-E_{\mathbf{m},\mathbf{K}-\mathbf{q}}^{\left(f\right)}$,  the interchannel coupling
\begin{align}
\label{eq:hoverlaps}\frac{h_{\mathbf{s}\mathbf{K}}^{\mathbf{nm}}\left(\mathbf{q}\right)}{\sqrt{N^3v}}&=\int d\mathbf{x}d\mathbf{y}\left[\phi^{\left(f\right)}_{\mathbf{n}\mathbf{q}}\left(\mathbf{x}\right)\phi_{\mathbf{m},\mathbf{K}-\mathbf{q}}^{\left(f\right)}\left(\mathbf{y}\right)\right]^{\star}\alpha\left(\mathbf{r}\right)\phi_{\mathbf{s}\mathbf{K}}^{\left(b\right)}\left(\mathbf{R}\right)\, ,
\end{align}
and $v$, the volume of a unit cell.  Note that we have affixed the subscript $\mathbf{K}$ to the energy eigenvalue to denote that it is the eigenvalue for fixed total quasimomentum.  Taking the continuum limit allows us to make the replacement 
\begin{align}
\label{eq:continuumlimit}\sum_{\mathbf{q}}&\to \frac{N^3}{v_{\mathrm{BZ}}}\int_{\mathrm{BZ}}\, ,
\end{align}
where $v_{\mathrm{BZ}}$ is the volume of the BZ.  The equations (\ref{eq:formereqn}-\ref{eq:lattereqn}) become
\begin{align}
\label{eq:firstofthetwo}[E_{\mathbf{K}}-E_{\mathbf{nm}}^{\mathbf{K}}\left(\mathbf{q}\right)&]\varphi_{\mathbf{nm}}^{\mathbf{K}}\left(\mathbf{q}\right)=\\
\nonumber &\frac{g}{\sqrt{v}}\sum_{\mathbf{s}}h_{\mathbf{s}\mathbf{K}}^{\mathbf{nm}}\left(\mathbf{q}\right)\Upsilon_{\mathbf{s}}^{\mathbf{K}}\, ,\\
\label{eq:secondofthetwo}[E_{\mathbf{K}}-h\nu_0-&E^{\left(b\right)}_{\mathbf{s}\mathbf{K}}]\Upsilon_{\mathbf{s}}^{\mathbf{K}}=\\
\nonumber &\frac{g}{\sqrt{v}}\sum_{\mathbf{n}\mathbf{m}}\int_{\mathrm{BZ}}\frac{d\mathbf{q}}{v_{\mathrm{BZ}}}{h_{\mathbf{s}\mathbf{K}}^{\mathbf{nm}}}^{\star}\left(\mathbf{q}\right)\varphi_{\mathbf{nm}}^{\mathbf{K}}\left(\mathbf{q}\right)\, ,
\end{align}
Formally solving Eq.~\eqref{eq:firstofthetwo} by applying $\left(E-\hat{H}_0\left(\mathbf{x}\right)-\hat{H}_0\left(\mathbf{y}\right)+i\eta\right)^{-1}$ with $\eta$ a positive infinitesimal and inserting into Eq.~\eqref{eq:secondofthetwo} gives 
\begin{align}
\label{eq:unrenormalized} [E_{\mathbf{K}}-\nu_0-&{E}^{\left(b\right)}_{\mathbf{s}\mathbf{K}}]\Upsilon_{\mathbf{s}}^{\mathbf{K}}=\\
\nonumber &\frac{g^2}{v}\sum_{\mathbf{t}}\left[\int_{\mathrm{BZ}}\frac{d\mathbf{q}}{v_{\mathrm{BZ}}}\sum_{\mathbf{n}\mathbf{m}}\frac{h_{\mathbf{s}\mathbf{K}}^{\mathbf{nm}}\left(\mathbf{q}\right) {h_{\mathbf{t}\mathbf{K}}^{\mathbf{nm}}}^{\star}\left(\mathbf{q}\right)}{E_{\mathbf{K}}-E_{\mathbf{nm}}^{\mathbf{K}}\left(\mathbf{q}\right)+i\eta}\right]\Upsilon_{\mathbf{t}}^{\mathbf{K}}\, .
\end{align}
The right-hand side of Eq.~\eqref{eq:unrenormalized} diverges in the limit $\Lambda\to \infty$, as is well known for two-channel theories involving a point-like boson~\cite{Kokkelmans_Milstein_02}.  We remove this divergence through renormalization, replacing the divergent bare detuning $h\nu_0$ with the physical detuning $h\nu$ by subtracting off the divergent part of the bare detuning~\cite{Kokkelmans_Milstein_02,HollandProceedings,HMV}.  Specifically, 
\begin{align}
\label{eq:NLEE}&\left[E_{\mathbf{K}}-h\nu-{E}_{\mathbf{s}\mathbf{K}}^{\left(b\right)}\right]\Upsilon_{\mathbf{s}}^{\mathbf{K}}=\frac{g^2}{v}\sum_{\mathbf{t}}\chi_{\mathbf{st}}^{\mathbf{K}}\left(E_{\mathbf{K}}\right)\Upsilon_{\mathbf{t}}^{\mathbf{K}}\, ,\\
\label{eq:chidef}&\chi_{\mathbf{st}}^{\mathbf{K}}\left(E_{\mathbf{K}}\right)\equiv\int_{\mathrm{BZ}}\frac{d\mathbf{q}}{v_{\mathrm{BZ}}}\sum_{\mathbf{n}\mathbf{m}}\frac{h_{\mathbf{s}\mathbf{K}}^{\mathbf{nm}}\left(\mathbf{q}\right) {h_{\mathbf{t}\mathbf{K}}^{\mathbf{nm}}}^{\star}\left(\mathbf{q}\right)}{E_{\mathbf{K}}-E_{\mathbf{nm}}^{\mathbf{K}}\left(\mathbf{q}\right)+i\eta}-\bar{\chi}_{\mathbf{st}}^{\mathbf{K}}\, ,\\
\label{eq:barchi}&\bar{\chi}_{\mathbf{st}}^{\mathbf{K}}\equiv\int_{\mathrm{BZ}}\frac{d\mathbf{q}}{v_{\mathrm{BZ}}}\sum_{\mathbf{n}\mathbf{m}}\frac{\bar{h}_{\mathbf{s}\mathbf{K}}^{\mathbf{nm}}\left(\mathbf{q}\right) {\bar{h}_{\mathbf{t}\mathbf{K}}^{\mathbf{nm\star}}}\left(\mathbf{q}\right)}{-\bar{E}_{\mathbf {nm}}^{\mathbf{K}}\left(\mathbf{q}\right)}\, ,
\end{align}
where the bars indicate that these quantities are computed in the absence of an optical lattice.  The renormalization matrix $\bar{\chi}_{\mathbf{st}}^{\mathbf{K}}$ is a diagonal matrix, as will be shown explicitly in Sec.~\ref{sec:properreg}.  The divergent parts of the two terms in Eq.~\eqref{eq:chidef} cancel each other and the limit $\Lambda\to \infty$ is finite.  

The renormalization prescription Eq.~\eqref{eq:chidef} is similar the one in free space, Eq.~\eqref{eq:freerenorm}.  There are two key differences.    The first is that the renormalization is not a scalar function as in free space, but rather a matrix indexed by closed channel band indices.  The matrix element $\bar{\chi}^{\mathbf{K}}_{\mathbf{ss}}$ is the renormalization of the detuning of a closed channel molecule in band $\mathbf{s}$ due to effective coupling through all bands of the open channel.  The second difference is that, rather than imposing a high-momentum cutoff as in free space, we perform the regularization through the function $\alpha(\mathbf{r})$  The two procedures are the same if $\alpha(\mathbf{r})$ is chosen to be the Fourier transform of a Heaviside step function in the radial momentum.  However, radially symmetric regularizations are inappropriate for the lattice renormalization due to a symmetry mismatch with the lattice.  A better approach is to take a regularization function $\alpha(\mathbf{r})$ which is the Fourier transform of a function with the symmetry of the reciprocal-space BZ.  Our specific regularization and the manner in which the limit $\Lambda\to\infty$ is taken are discussed in Sec.~\ref{sec:properreg}.  The computation of the overlaps $h_{\mathbf{sK}}^{\mathbf{nm}}\left(\mathbf{q}\right)$ is discussed in detail in Sec.~\ref{sec:MatrixElem}, and efficient numerical methods for performing the BZ integration in Eq.~\eqref{eq:chidef} and solving Eq.~\eqref{eq:NLEE} are presented in Appendices \ref{sec:GMA} and \ref{sec:NLEEsoln}, respectively.

The equation for the bound states, Eq.~\eqref{eq:NLEE}, is an eigenvalue problem which is nonlinear in the energy eigenvalue $E_{\mathbf{K}}$.  The eigenvectors $\left\{\boldsymbol{\Upsilon}^{\mathbf{K}}_{\alpha}\right\}$ satisfying this equation define the closed channel portion of the wavefunction through Eq.~\eqref{eq:molexpansion}.  The open channel portion can be found using
\begin{align}
\varphi_{\mathbf{nm}}^{\mathbf{K};\alpha}\left(\mathbf{q}\right)&=\frac{g}{\sqrt{v}}\sum_{\mathbf{s}}\frac{h_{\mathbf{sK}}^{\mathbf{nm}}\left(\mathbf{q}\right)\Upsilon_{\mathbf{s}\alpha}^{\mathbf{K}}}{E_{\mathbf{K}}^{\alpha}-E_{\mathbf{nm}}^{\mathbf{K}}\left(\mathbf{q}\right)+i\eta}
\end{align} 
together with Eq.~\eqref{eq:openchannel}.  Here $\alpha$ is an index denoting independent solutions of Eq.~\eqref{eq:NLEE} for fixed $g$ and $\nu$.  Hence, the two-particle solutions of Eq.~\eqref{eq:NLEE} may be written as
\begin{align}
\label{eq:tpstate}&\Psi_{\mathbf{K}\alpha}(\mathbf{x},\mathbf{y})=\frac{1}{\mathcal{N}_{\mathbf{K}\alpha}}\Big[\sum_{\mathbf{s}}\Upsilon^{\mathbf{K}}_{\mathbf{s}\alpha}\phi^{\left(b\right)}_{\mathbf{sK}}(\mathbf{x})\kappa(\mathbf{x}-\mathbf{y})\\
\nonumber&+\frac{g}{\sqrt{N^3v}}\sum_{\mathbf{nms};\mathbf{q}}\frac{\Upsilon_{\mathbf{s}\alpha}^{\mathbf{K}}h_{\mathbf{sK}}^{\mathbf{nm}}(\mathbf{q})\phi^{\left(f\right)}_{\mathbf{nq}}(\mathbf{x})\phi^{\left(f\right)}_{\mathbf{m},\mathbf{K}-\mathbf{q}}(\mathbf{y})}{E_{\mathbf{K}}^{\alpha}-E_{\mathbf{nm}}^{\mathbf{K}}(\mathbf{q})}\Big]\, ,
\end{align}
where $\mathcal{N}_{\mathbf{K}\alpha}$ is a normalizing factor and $\kappa(\mathbf{x}-\mathbf{y})$ denotes a relative wavefunction for the closed channel which has characteristic volume $v/\Lambda^3$.  At the end of the computation, we allow $\Lambda\to \infty$, see Sec.~\ref{sec:properreg}, and so the purpose of $\kappa$ is to remind that the closed channel contributes only at the center of mass.  The normalization coefficient is
\begin{align}
\mathcal{N}_{\mathbf{K}\alpha}^2&=1-(\frac{g}{E_R\sqrt{v}})^2\boldsymbol{\Upsilon}^{\mathbf{K}}_{\alpha}\cdot\frac{\partial \chi^{\mathbf{K}}(E_{\mathbf{K}}^{\alpha}/E_R)}{\partial E_{\mathbf{K}}}\cdot\boldsymbol{\Upsilon}^{\mathbf{K}}_{\alpha}\, .
\end{align}

In the present work we specialize to the case of a simple cubic optical lattice $V_{\mathrm{latt}}\left(\mathbf{x}\right)=V\sum_{i=x,y,z}\sin^2\left(k_li\right)$ with $k_l=\pi/a$, $a$ the lattice spacing.  Thus, the volume of the unit cells in real and reciprocal space are $v=a^3$ and $v_{\mathrm{BZ}}=8\pi^3/v$, respectively, and the natural unit of energy is the recoil energy $E_R=\hbar^2k_l^2/2m$.  The parameterization of the interchannel coupling and the scattering length in natural units are
\begin{align}
\label{eq:natunits1}\tilde{g}&\equiv \frac{g}{E_Ra^{3/2}}=\frac{4}{\pi^{3/2}}\sqrt{\frac{a}{r_B}}\,,\\
\label{eq:natunits2}\tilde{a}_s&\equiv \frac{a_s}{a}=-\frac{\pi}{8}\frac{g^2}{E_R^2a^3}\frac{E_R}{h\nu}\,.
\end{align}
Additionally, the zero of energy is chosen to be the energy of two particles at zero quasimomentum in the lowest Bloch band, $E_{\mathbf{11}}^{\mathbf{0}}\left(\mathbf{0}\right)$.

\subsection{Computation and symmetries of the single-particle basis}
\label{sec:spandsymm}
In this section we discuss how to efficiently solve the single-particle Hamiltonians Eqs.~(\ref{eq:Bloch1}-\ref{eq:Bloch2}) and discuss how the single-particle eigenfunctions transform under the point group symmetries of the lattice.  The simple cubic lattice is separable in the sense that the lattice potential is a sum of 1D lattice potentials
\begin{align}
\label{eq:Lattdef}V_{\mathrm{latt}}\left(\mathbf{x}\right)=V\sum_{i=x,y,z}\sin^2\left(k_li\right)\, .
\end{align}
Hence, the 3D Bloch functions $\phi_{\mathbf{n}\mathbf{q}}\left(\mathbf{x}\right)$ which form a complete eigenbasis of the single-particle Hamiltonian may be written as products of 1D Bloch functions $\phi_{nq}\left(x\right)$ which satisfy
\begin{align}
\left[\frac{\hat{p}^2}{2m}+V\sin^2\left(\frac{\pi x}{a}\right)\right]\phi_{nq}\left(x\right)&=E_{nq}\phi_{nq}\left(x\right)\, .
\end{align}
Here, $n$ denotes a one-dimensional band index and $q$ a one-dimensional quasimomentum.  We compute the 1D Bloch functions numerically by writing 
\begin{align}
\label{eq:unitperiodic} \phi_{nq}\left(x\right)=e^{iqx}u_{nq}\left(x\right)/\sqrt{Na}
\end{align}
where $u_{nq}\left(x\right)=u_{nq}\left(x+a\right)$ are the eigenstates of the operator
\begin{align}
\hat{H}_q&=\frac{\left(\hat{p}+q\right)^2}{2m}+V\sin^2\left(\frac{\pi x}{a}\right)\, .
\end{align}
Because of their periodicity, the functions $u_{nq}\left(x\right)$ may be expanded in a Fourier series as
\begin{align}
\label{eq:uFourier}u_{nq}\left(x\right)&=\lim_{l\to\infty}\sum_{j=-l}^{l}c_{nq}^j e^{2\pi i j x/a}\, .
\end{align}
The real vector with elements $\{{c}_{nq}^j\}$ satisfies the eigenvalue equation $\sum_{j'}H_{jj'}^qc_{n,q}^{j'}=E_{nq}c_{n,q}^{j}$ where $H^q_{jj'}=\left(2j+q/k_l\right)^2/E_R$ for $j=j'$, $-V/4$ for $\left|j-j'\right|=1$, and $0$ otherwise.  The fact that the matrix $H^q$ is tridiagonal stems from the fact that the lattice potential, Eq.~\eqref{eq:Lattdef}, contains only three Fourier components.  Other lattices can be treated in an analogous manner.  Convergence is achieved to machine precision even for high band numbers $\sim 20$ by taking a finite Fourier cutoff $l$ on the order of a few tens.

In addition to translational symmetry, the single particle Hamiltonian resulting from the simple cubic lattice is invariant under inversions of any set of Cartesian coordinates.  Here, we discuss the transformations of the single-particle functions $\phi_{\mathbf{nq}}(\mathbf{x})$ in detail, as it provides a useful means to classify the interacting two-particle solutions, Eq.~\eqref{eq:tpstate}, see Sec.~\ref{sec:Wannier}.  Additionally, proper care of the lattice symmetries is important to ensuring that the parameters in the effective many-body model transform sensibly under point group transformations.  We will denote the operator performing a spatial inversion as $\hat{\theta}_{\mathcal{R}}$, where $\mathcal{R}$ is the set of Cartesian dimensions which are to be inverted.  Additionally, an inverted coordinate will be denoted with a prime.  That is, $\hat{\theta}_{\mathcal{R}}\hat{\mathbf{x}}=\hat{\mathbf{x}}'$ where $x_j'=-x_j$ for all $j\in\mathcal{R}$ and $x_j'=x_j$ otherwise.  When the Hamiltonian is time-reversal invariant the presence of an inversion symmetry generated by $\theta_{\mathcal{R}}$ implies that the Bloch functions of the simple cubic lattice transform as
\begin{align}
\label{eq:inversion}\phi_{\mathbf{n}\mathbf{q}}\left(\mathbf{x}'\right)&=\prod_{j\in\mathcal{R}}\left(-1\right)^{n_j+1} \phi_{\mathbf{n}\mathbf{q}'}\left(\mathbf{x}\right)\, .
\end{align}
At the extremal points of the BZ, $q=0$ and $q=\pi/a$, translations and inversions commute and so the associated Bloch functions have well-defined parity with respect to inversion.  At all other points in the BZ, translation and inversion do not commute, but the inversion operator connects degenerate states at different points in the BZ up to a phase.  From Eq.~\eqref{eq:inversion}, we see that $u_{nq}\left(-x\right)=\left(-1\right)^{n+1}u_{n,-q}\left(x\right)$.  For the extremal points in the BZ this implies that the vectors $\mathbf{c}_{nq}$ must have definite parity about $j=0$ and $j=1$, respectively, see Eq.~\eqref{eq:uFourier}.  When computing the vectors $\mathbf{c}_{nq}$ numerically the even and odd parity states become quasi-degenerate for high bands where the band gap becomes smaller than machine precision and the numerical eigenvectors will not automatically have definite parity.  However, it is always possible to sort the eigenstates according to their parity to maintain the relation $u_{nq}\left(-x\right)=\left(-1\right)^{n+1}u_{n,-q}\left(x\right)$.  There is still an additional global phase ambiguity between Bloch functions with different values of the quasimomentum $\mathbf{q}$.  We fix this phase ambiguity by requiring that the Bloch functions are smoothly varying functions of $\mathbf{q}$.  This phase prescription gives rise to maximally localized Wannier functions~\cite{Kohn_59}.

The inversion relationship Eq.~\eqref{eq:inversion} implies that the matrix elements of the inter-channel coupling transform as
\begin{align}
h_{\mathbf{s}\mathbf{K}'}^{\mathbf{nm}}\left(\mathbf{q}'\right)&=\prod_{j\in\mathcal{R}}\left(-1\right)^{n_j+m_j+s_j+1}h_{\mathbf{s}\mathbf{K}}^{\mathbf{nm}}\left(\mathbf{q}\right)\, ,
\end{align}
and so
\begin{align}
\chi_{\mathbf{st}}^{\mathbf{K}'}\left(E_{\mathbf{K}}\right)&=\prod_{j\in\mathcal{R}}\left(-1\right)^{s_j+t_j}\chi_{\mathbf{st}}^{\mathbf{K}}\left(E_{\mathbf{K}}\right)\, .
\end{align}
At extremal points of the total quasimomentum BZ, $\mathbf{K}=0$ and $-k_l\left(1,1,1\right)$, the inversion properties of $\chi^{\mathbf{K}}$ imply that only molecular bands which transform identically under complete inversions mix.  Numerically, we find that the mixing between bands even for arbitrary $\mathbf{K}$ is generally small for energies near the lowest open channel band except at isolated points where eigenvalues cross.  Inversion leaves the eigenvalues of $\chi^{\mathbf{K}}\left(E_{\mathbf{K}}\right)$ invariant, but the eigenvectors transform as
\begin{align}
\Upsilon^{\mathbf{K}'}_{\mathbf{s}\alpha}&=P_{\alpha}\prod_{j\in\mathcal{R}}\left(-1\right)^{s_j+1}\Upsilon^{\mathbf{K}}_{\mathbf{s}\alpha}\, ,
\end{align}
where $P_{\alpha}$ denotes a unimodular phase.  Because the interaction is invariant under all elements of the inversion group we can classify the eigenvectors $\boldsymbol{\Upsilon}^{\mathbf{K}}_{\alpha}$ according to their inversion properties~\cite{wall_carr_12}.  Accordingly, we choose $P_{\alpha}=\prod_{j\in\mathcal{R}}\left(-1\right)^{s_j+1}$, and require that $\Upsilon^{\mathbf{K}}_{\mathbf{s}\alpha}$ be a smooth function of $\mathbf{K}$ otherwise.  This fixing of arbitrary phases implies that the interacting two-particle solutions, Eq.~\eqref{eq:tpstate}, are also smooth functions of $\mathbf{K}$.  Importantly, this convention also ensures that the Wannier functions have their expected behavior with respect to the symmetries of the lattice, see Sec.~\ref{sec:Wannier}.

\subsection{Matrix elements of the inter-channel coupling}
\label{sec:MatrixElem}
We now focus on the computation of the matrix elements of the inter-channel coupling, Eq.~\eqref{eq:hoverlaps}.  We take as our regularization
\begin{align}
\label{eq:alphadef} \alpha\left(\mathbf{r}\right)&=\int_{v\left(\Lambda\right)}\frac{d\mathbf{k}}{\left(2\pi\right)^3}e^{-i\mathbf{k}\cdot\mathbf{r}}
\end{align}
where $v\left(\Lambda\right)=v_{\mathrm{BZ}}\Lambda^3$ and $\Lambda$ is a positive integer.  This regularization is similar to the momentum-shell regularization with cutoff $K^{\star}$ in free space, see Eq.~\eqref{eq:freerenorm}.  However, rather than choosing a spherically symmetric cutoff, we choose the cutoff function to have the same symmetry as the BZ.  The integral Eq.~\eqref{eq:alphadef} may be explicitly evaluated to yield
\begin{align}
\label{eq:regular}\alpha\left(\mathbf{r}\right)&=\prod_{\nu=x,y,z}\frac{\sin\left(\Lambda \pi r_{\nu}\right)}{\pi r_{\nu}}\, ,
\end{align}
which is separable along principal axes and an even function.  For the simple cubic lattice, this implies that the 3D overlaps Eq.~\eqref{eq:hoverlaps} are products of 1D overlaps.  Objects in 1D will be distinguished from their 3D counterparts by not having bold symbols as arguments.  We note that each overlap as written is dimensionless.

The 1D overlaps take the form
\begin{align}
\nonumber h_{sK}^{nm}\left(q\right)=&\sqrt{N}\int_0^{N}dx_1\int_0^{N}dx_2 \left[\phi^{\left(f\right)}_{nq}\left(x_1\right)\phi^{\left(f\right)}_{m,K-q}\left(x_2\right)\right]^{\star}\\
&\times \phi_{sK}^{\left(b\right)}\left(\frac{x_1+x_2}{2}\right)\alpha\left(x_1-x_2\right)\, .
\end{align}
The highly oscillatory nature of both the Bloch functions and the regularization makes the numerical evaluation of this integral by quadrature computationally expensive.  Rather, we seek to find an analytic expression for this integral in terms of the Fourier components of the Bloch functions, see Eq.~\eqref{eq:uFourier}.

We begin by changing integration variables to $\xi=\left(x_1+x_2\right)/2$, $\eta=\left(x_1-x_2\right)/2$ with Jacobian 2.  The region of integration is bounded by the parametric $\left(\eta,\xi\right)$ curves as $\left(x,x\right)$, $x\in\left[0,\frac{N}{2}\right]$; $\left(x,-x\right)$, $x\in\left[0,-\frac{N}{2}\right]$; $\left(x,N-x\right)$, $x\in\left[\frac{N}{2},0\right]$; and $\left(x,N+x\right)$, $\left[-\frac{N}{2},0\right]$.  The integral becomes
\begin{align}
\nonumber \frac{h_{sK}^{nm}\left(q\right)}{2\sqrt{N}}=& \sum_{p=\left\{-1,1\right\}} p \int_0^{pN/2}d\eta \alpha\left(2\eta\right)\int_{p\eta}^{N-p\eta}d\xi \phi_{sK}^{\left(b\right)}\left(\xi\right)\\
&\times \phi_{nq}^{\left(f\right)\star}\left(\xi+\eta\right)\phi_{m,K-q}^{\left(f\right)\star}\left(\xi-\eta\right)
\end{align}
Expanding the Bloch functions as in Eq.~\eqref{eq:unitperiodic}, we have
\begin{align}
\nonumber &\frac{N h_{sK}^{nm}\left(q\right)}{2}=\sum_{p=\left\{-1,1\right\}}  p \int_0^{pN/2}d\eta e^{-i\eta\left(2q-K-2\pi \Gamma\right)}\alpha\left(2\eta\right)\\
\label{eq:hshiftdef}&\times \int_{p\eta}^{N-p\eta} d\xi e^{-i 2\pi \Gamma \xi} u_{sK}\left(\xi\right) u_{nq}^{\star}\left(\xi+\eta\right)u_{m,K-q}^{\star}\left(\xi-\eta\right)\, ,
\end{align}
where $\Gamma$ is an integer such that $2\pi \Gamma$ shifts $K-q$ to lie in the BZ.  If we now expand the functions $u$ in terms of their Fourier components according to Eq.~\eqref{eq:uFourier}, we find
\begin{align}
\nonumber &\frac{N h_{sK}^{nm}\left(q\right)}{2}=\lim_{l\to\infty}\sum_{j,j',j''=-l}^{l} \sum_{p=\left\{-1,1\right\}}  p c_{nq}^{j}c_{m,K-q}^{j'}c_{sK}^{j''}\\
\nonumber &\times  \int_0^{pN/2}d\eta e^{-i\eta\left[\left(2q-K\right)+2\pi\left(j-j'-\Gamma\right)\right]}\alpha\left(2\eta\right)\\
&\times \int_{p\eta}^{N-p\eta} d\xi e^{-i2 \pi \xi\left[\Gamma+j+j'-j''\right]}\, .
\end{align}
Changing integration variables to $\eta=p\eta$ and using the fact that $\alpha\left(\eta\right)$ is an even function yields
\begin{align}
\nonumber &\frac{N h_{sK}^{nm}\left(q\right)}{4}=\lim_{l\to\infty}\sum_{j,j',j''=-l}^{l}c_{nq}^{j}c_{m,K-q}^{j'}c_{sK}^{j''}\\
 &\times  \int_0^{N/2}d\eta \cos\left(2\pi z\eta\right)\alpha\left(2\eta\right) \int_{\eta}^{N-\eta} d\xi e^{i2\pi \xi t}\, ,
\end{align}
where
\begin{align}
\label{eq:zequiv}z&\equiv \frac{2q-K}{2\pi}+\left(j-j'-\Gamma\right)\, ,\\
\label{eq:tequiv} t&\equiv j''-j-j'-\Gamma\, .
\end{align}

We now focus on the $\xi$ integral:
\begin{align}
\nonumber \int_{\eta}^{N-\eta}d\xi e^{2\pi i\xi t}&=\left\{\begin{array}{c} \frac{1}{2\pi i t}{\left(e^{2\pi i\left(N-\eta\right)t}-e^{2 \pi i \eta t}\right)}\;\;\;\; t \ne 0\\
 N-2\eta\;\;\;\mbox{otherwise}\end{array}\right.
\end{align}
 Noting that $t$ is always an integer, we have
\begin{align}
 \int_{\eta}^{N-\eta}d\xi e^{2\pi i\xi t} &=N\delta_{t,0}-\frac{\sin 2\pi \eta t}{\pi t}\, .
\end{align}
Hence, the overlaps may be written
\begin{align}
\nonumber &h_{sK}^{nm}\left(q\right)=\lim_{l\to\infty}2\sum_{j,j'=-l}^lc_{n,q}^jc_{m,K-q}^{j'}c_{s,K}^{j+j'+\Gamma} I_1(N,\Lambda,z)\\
&-\lim_{l\to\infty}\frac{2}{N}\sum_{j,j',j''=-l}^{l}c_{n,q}^jc_{m,K-q}^{j'}c_{s,K}^{j''}I_2(N,\Lambda,z,t)\, ,
\end{align}
where
\begin{align}
I_1(N,\Lambda,z)&=\int_0^{N/2}d\eta \frac{\sin\left(2\pi \Lambda \eta\right)}{\pi \eta}\cos\left(2\pi \eta z\right)\, ,\\
I_2(N,\Lambda,z,t)&=\int_0^{N/2}d\eta \frac{\sin \left(2\pi \Lambda \eta \right)}{\pi \eta}\cos\left[2\pi \eta z\right]\frac{\sin 2\pi \eta t}{\pi t}\, .
\end{align}
In these integrals we have substituted the actual form of the regularization, Eq.~\eqref{eq:regular}.  The first integral is
\begin{align}
I_1(N,\Lambda,z)&=\frac{1}{2\pi}\left[\mathrm{Si}\left(N\pi\left(\Lambda+z\right)\right)+\mathrm{Si}\left(N\pi\left(\Lambda-z\right)\right)\right]
\end{align}
where the sine integral $\mathrm{Si}$ has the definition
\begin{align}
\mathrm{Si}\left(y\right)&=\int_0^ydx\frac{\sin x}{x}\, .
\end{align}
For the second integral, we first note that
\begin{align}
\lim_{t\to 0}I_2(N,\Lambda,z,t)&=\frac{\Lambda\left(\cos\left(N\pi\Lambda\right)\cos\left(N\pi z\right)-1\right)}{\pi^2\left(z^2-\Lambda^2\right)}\, .
\end{align}
As we anticipate taking the limit $N\to \infty$, we may choose that $N$ is even without loss of generality and so
\begin{align}
\lim_{t\to 0} I_2(N,\Lambda,z,t)&=\frac{\Lambda\left(\cos\left(N\pi z\right)-1\right)}{\pi^2\left(z^2-\Lambda^2\right)}\, ,
\end{align}
which vanishes as $z\to \Lambda$.  For $t\ne 0$, we have
\begin{align}
\nonumber 4\pi^2 t I_2(N,\Lambda,z,t)&=\mathcal{C}\left(N,t-z-\Lambda\right)+\mathcal{C}\left(N,t+z-\Lambda\right)\\
&-\mathcal{C}\left(N,t-z+\Lambda\right)-\mathcal{C}\left(N,t+z+\Lambda\right)\, ,
\end{align}
where
\begin{align}
\mathcal{C}\left(N,x\right)&\equiv \mathrm{Ci}\left(N \pi x\right)-\log x\, ,\\
\mathrm{Ci}\left(x\right)&\equiv\gamma+\log x+\int_0^xdt\frac{\cos t-1}{t}\, .
\end{align}
Here, $\gamma\approx 0.5772$ is the Euler-Mascheroni constant and $\mathrm{Ci}\left(x\right)$ is the cosine integral.  The function $\mathcal{C}\left(N,x\right)$ is even in $x$, and so we may take the absolute values of the $x$ arguments without loss of generality.  Furthermore, while the both the cosine integral and the logarithm diverge as $x\to 0$, $\lim_{x\to 0}\mathcal{C}\left(N,x\right)=\gamma+\log N\pi$.

In summary, we have
\begin{widetext}
\begin{align}
\label{eq:hfullN}h_{sK}^{nm}\left(q\right)&=\frac{1}{\pi}\sum_{jj'}c_{n,q}^{j}c_{m,K-q}^{j'}c_{sK}^{j+j'+\Gamma}\left[\mathrm{Si}\left(N\pi\left(\Lambda+z\right)\right)+\mathrm{Si}\left(N\pi\left(\Lambda-z\right)\right)+\frac{2\Lambda\left(\cos\left(N\pi z\right)-1\right)}{N\pi\left(\Lambda^2-z^2\right)}\right]\\
\nonumber &-\sum_{jj'j''}'c_{n,q}^{j}c_{m,K-q}^{j'}c_{sK}^{j''}\frac{1}{2\pi^2 Nt}\left[\mathcal{C}(N,\left|t-z-\Lambda\right|)+\mathcal{C}(N,\left|t+z-\Lambda\right|)-\mathcal{C}(N,\left|t-z+\Lambda\right|)-\mathcal{C}(N,\left|t+z+\Lambda\right|)\right]\, ,
\end{align}
\end{widetext}
where the prime on the latter summation indicates that terms where $t=0$ are excluded.  Let us now consider taking the limit of an infinite number of unit cells, $N\to \infty$.  For the primed summation, we have that $\lim_{x\to\infty}\mathrm{Ci}\left(x\right)=0$ and the logarithms are independent of $N$, so this entire summation vanishes as $N\to\infty$, $x\ne 0$.  The only other possibility is that one of the terms in the absolute values is identically zero, but in that case $\lim_{x\to 0}\mathcal{C}\left(N,x\right)=\gamma+\log N\pi$.  All such terms in the primed summation vanish as $N\to \infty$ because of the factor of $N$ in the denominator.  Hence, the primed summation has no contribution in the limit of an infinite lattice.

For the first summation in Eq.~\eqref{eq:hfullN}, we note that the last term in brackets vanishes as $N\to \infty$, as $z$ is real.  When $z=\Lambda$ this quantity vanishes identically for any $N$, as $\Lambda$ is defined to be an integer.  Hence, the only terms that remain are the sine integrals.  We note that $\lim_{x\to \infty}\mathrm{Si}\left(x\right)=\pi/2$, but also that $\mathrm{Si}$ is an odd function.  Thus,
\begin{align}
\nonumber \lim_{N\to \infty}\frac{1}{\pi}\Big[\mathrm{Si}\left(N\pi\left(\Lambda+z\right)\right)+&\mathrm{Si}\left(N\pi\left(\Lambda-z\right)\right)\Big]\\
&=\left\{\begin{array}{cc} 1\;\;\; &-\Lambda<z<\Lambda\\ \frac{1}{2}\;\;\; &\left|z\right|=\Lambda\\ 0\;\;\;&\mathrm{otherwise}\end{array}\right.\, ,
\end{align}
and so
\begin{align}
\label{eq:hover}\lim_{N\to \infty} h_{sK}^{nm}\left(q\right)=&\sum_{jj'}c_{n,q}^{j}c_{m,K-q}^{j'}c_{sK}^{j+j'+\Gamma}\\
\nonumber &\times \mathrm{rect}\left[\frac{2q-K+2\pi\left(j-j'-\Gamma\right)}{2\pi \Lambda}\right]\, ,
\end{align}
where $\mathrm{rect}\left(x\right)$ is the rectangle function.  In the limit of an infinite lattice, the effect of the regularization is to cut off the contribution of the overlap in the shifted relative Fourier space defined by $z$, see Eq.~\eqref{eq:zequiv}, with a rectangle function of width $2\Lambda$.  For finite $N$, the overlaps Eq.~\eqref{eq:hfullN} display a pronounced Gibbs phenomenon which make integration difficult near $z=\Lambda$.  By taking the $N\to\infty$ limit the function becomes much better behaved.  Additionally, taking $N\to \infty$ allows us to consistently make the claim that the quasimomentum $q$ is a continuous variable and so replace all sums by integrals according to Eq.~\eqref{eq:continuumlimit} in Sec.~\ref{sec:FRHPRA:fewbody}.

\subsection{Regularization of the theory and evaluation of the renormalization at $\mathbf{K}=0$}
\label{sec:properreg}
In this section we use the results of Sec.~\ref{sec:MatrixElem} to find an expression for the renormalization matrix, Eq.~\eqref{eq:barchi}, at zero total quasimomentum.  In addition to the theoretical appeal of being able to compute the renormalization analytically, this procedure also sheds light on the proper regularization procedure.  Let us first define a shell summation over bands with shell parameter $S$, $\sum_{\mathbf{nm};S}$, as the summation over all $\mathbf{n}$ and $\mathbf{m}$ whose components are less than or equal to $S$ with at least one of the components being $S$.  To illustrate this notation, the shell summation for $S=1$ consists of $\mathbf{n}=(1,1,1)$, $\mathbf{m}=(1,1,1)$:
\begin{align}
\sum_{\mathbf{nm};1}&=\sum_{n_x=1}^{1}\sum_{n_y=1}^{1}\sum_{n_z=1}^{1}\sum_{m_x=1}^{1}\sum_{m_y=1}^{1}\sum_{m_z=1}^{1}\, .
\end{align}
The shell summation for $S=2$ is
\begin{align}
\nonumber \sum_{\mathbf{nm};2}=&\sum_{n_x=1}^{2}\sum_{n_y=1}^{2}\sum_{n_z=1}^{2}\sum_{m_x=1}^{2}\sum_{m_y=1}^{2}\sum_{m_z=1}^{2}\\
&-\sum_{n_x=1}^{1}\sum_{n_y=1}^{1}\sum_{n_z=1}^{1}\sum_{m_x=1}^{1}\sum_{m_y=1}^{1}\sum_{m_z=1}^{1}\, .
\end{align}
The subtracted summation ensures that at least of one the elements of $\mathbf{n}$ or $\mathbf{m}$ is 2, as required by our definition.  Shell summation provides a way to systematically add contributions from higher bands to a summation, as the summation over all bands with components less than or equal to $\ell$ is equal to the summation over all bands with components less than or equal to $\left(\ell-1\right)$ plus $\sum_{\mathbf{nm};\ell}$.  By induction, the summation over all bands with components less than or equal to $S$ is $\sum_{\ell=1}^{S}\sum_{\mathbf{nm};\ell}$.

To properly renormalize the theory and to include the effects of all higher bands one must take both the shell of open channel bands $S\to\infty$ and the regularization cutoff $\Lambda\to\infty$.  However, as shown in Ref.~\cite{Buechler_12}, the order of these limits is important, and the proper limiting procedure is first to take the shell parameter $S\to\infty$ with $\Lambda$ held fixed and then let $\Lambda\to\infty$.  The limit of infinite cutoff $\Lambda$ is taken using the asymptotic scaling relation~\cite{Buechler_12}
\begin{align}
\label{eq:FRHPRL:chiscaling}\left[\chi_{\mathbf{st}}^{\mathbf{K}}(E_{\mathbf{K}})\right](\Lambda)&=p_{\mathbf{st}}/\Lambda+\chi_{\mathbf{st}}^{\mathbf{K}}(E_{\mathbf{K}})\, ,
\end{align}
where $p_{\mathbf{st}}$ is the slope defining the scaling with $\Lambda$ and $\chi_{\mathbf{st}}^{\mathbf{K}}(E_{\mathbf{K}})$ is the value as $\Lambda\to\infty$.

Let us now consider computing the renormalization matrix, Eq.~\eqref{eq:barchi}, at $\mathbf{K}=0$.  Because the renormalization is computed in the absence of an optical lattice, the Fourier expansions of the single-particle states, Eq.~\eqref{eq:uFourier}, contain only a single plane wave, and the open channel energy becomes  $\bar{E}_{\mathbf{nm}}^{\mathbf{0}}\left(\mathbf{q}\right)=2E_R\mathbf{q}\cdot\mathbf{q}/\pi^2$.  The fact that all Fourier expansions contain a single plane wave implies that the overlaps $\bar{h}_{sK}^{nm}(q)$ with $s$ and $K$ fixed are nonzero only for a specific pair of open channel band indices $(n,m)$.  This in turn implies that the renormalization matrix $\bar{\chi}^{\mathbf{K}}$, defined in Eq.~\eqref{eq:barchi}, is diagonal.  The non-vanishing Fourier component for the lowest closed channel band with $\mathbf{s}=\left(1,1,1\right)$ is $\bar{c}_{\mathbf{10}}^{j}=\delta_{j,0}$.  Furthermore, the shift $\Gamma$ introduced in Eq.~\eqref{eq:hshiftdef} is $0$ when $\mathbf{K}=0$, and so
\begin{align}
\bar{h}_{\mathbf{10}}^{\mathbf{11}}\left(\mathbf{q}\right)&=\prod_{\nu=\left\{x,y,z\right\}}\left[\Theta\left(q_{\nu}+\Lambda\right)-\Theta\left(q_{\nu}-\Lambda\right)\right]\, ,
\end{align}
with $\Theta\left(x\right)$ the Heaviside function.  Thus, provided that $\Lambda\ge 1$, we have that the $S=1$ shell contribution to the $\mathbf{s}=\mathbf{1}$, $\mathbf{t}=\mathbf{1}$ element of the renormalization is
\begin{align}
\bar{\chi}_{\mathbf{11};S=1}^{\mathbf{0}}&=-\frac{1}{16 E_R}\int_{-1}^{1}dx\int_{-1}^1dy\int_{-1}^{1}dz\frac{1}{x^2+y^2+z^2}\, .
\end{align}
This integral cannot be evaluated analytically, but its numerical value is $\approx(-0.959265\pm 2\times 10^{-6}) /E_R$.  The evaluation of this integral and the error estimation were performed using the Genz-Malik algorithm, discussed in Appendix~\ref{sec:GMA}.

\begin{figure}[thbp]
\centerline{\includegraphics[width=0.9\columnwidth]{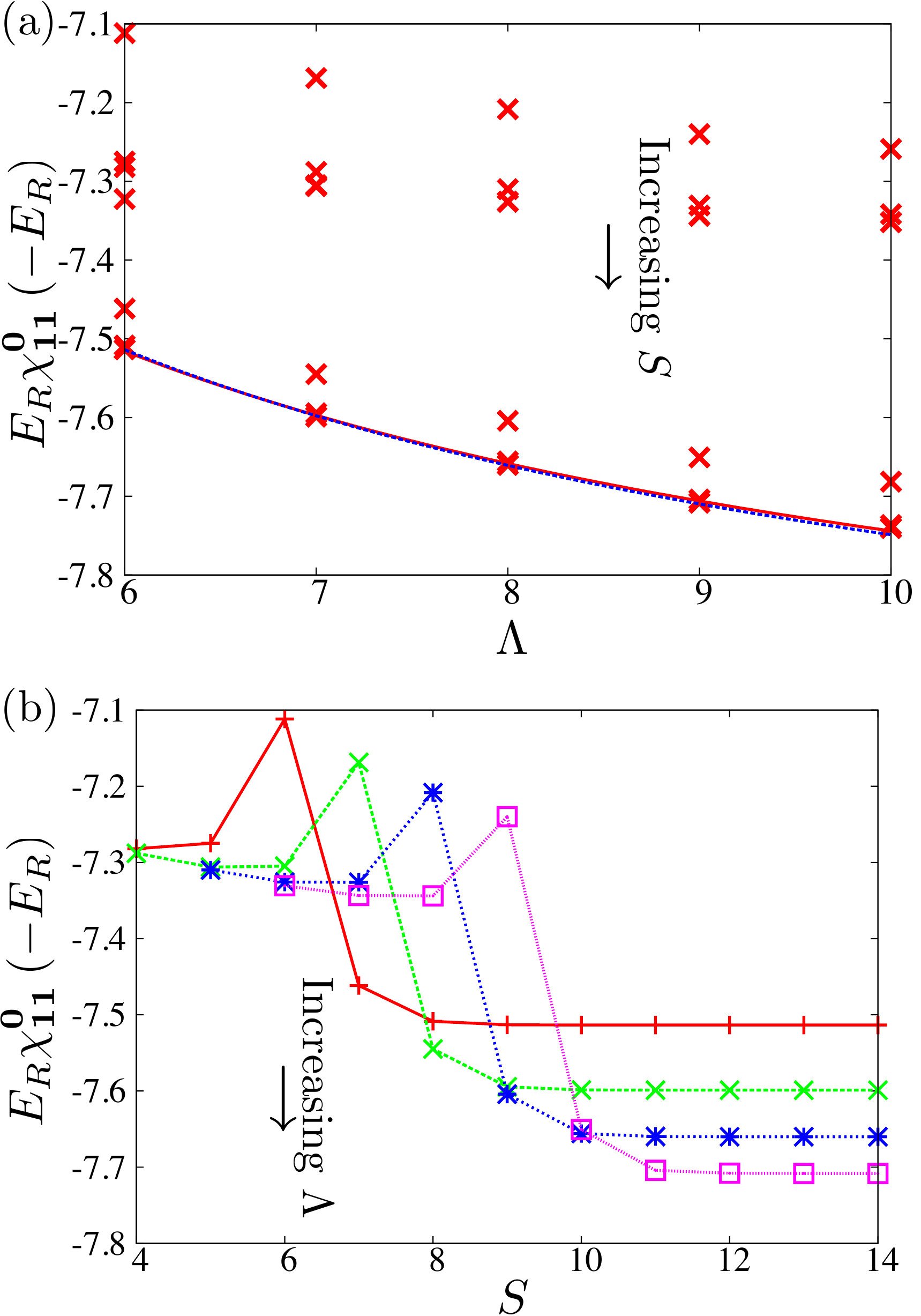}}
\caption{\label{fig:ScalinginLambdaandS} (Color online) \emph{Scaling of the effective lowest-band closed channel $T$-matrix $\chi^{\mathbf{K}}_{\mathbf{11}}$ with open channel band cutoff $S$ and regularization cutoff $\Lambda$.}  (a) The $T$-matrix element $\chi_{\mathbf{11}}^{\mathbf{0}}(E_{\mathbf{K}})$ computed at bound state energy $E_{\mathbf{K}}=-1E_R$ for a variety of open channel band cutoffs $S$ and regularization cutoffs $\Lambda$.  The open channel band cutoff $S$ is increasing towards the bottom of the figure.  The crosses are data and the solid lines are fits to the $S\to\infty$ scaling form Eq.~\eqref{eq:FRHPRL:chiscaling}, with the solid red line being a fit through the first three points and the dashed blue line being a fit through all five points.  The predictions from these two fits differ only by a few percent.  (b) The $T$-matrix element $\chi_{\mathbf{11}}^{\mathbf{0}}$ computed at $\Lambda=6$ (red pluses), $7$ (green crosses), $8$ (blue asterisks), and $9$ (purple boxes) for a variety of open channel band cutoffs $S$.  Near $S=\Lambda$ a marked change in behavior occurs followed by rapid convergence to the $S\to\infty$ limiting value.}
\end{figure}

Let us now consider contributions from the second shell, $S=2$.  From the constraint that only $j+j'=0$ terms contribute, see Eq.~\eqref{eq:hover}, we have that the only nonzero $\bar{h}_{sK}^{nm}\left(q\right)$ overlaps along each dimension have $n$ and $m$ either both in the lowest band or both in the first excited band.  Because of the shell summation constraint that at least one of the band indices is greater than 1, for each nonzero contribution we will have at least one pair of open channel band indices $\left(n,m\right)$ in the first excited band.  Now, integration over the energy of the excited band can be done in the extended zone scheme by integrating from $q=-2$ to $-1$ and then from $1$ to $2$.  The relevant integrals are
\begin{align}
\nonumber I_1=&-\frac{1}{16E_R}\left[\int_{-2}^{-1}dx+\int_{1}^{2}dx\right]\int_{-1}^1dy\\
\nonumber &\times \int_{-1}^{1}dz\frac{1}{x^2+y^2+z^2}\, ,\\
\nonumber I_2=&-\frac{1}{16E_R}\left[\int_{-2}^{-1}dx+\int_{1}^{2}dx\right]\left[\int_{-2}^{-1}dy+\int_{1}^{2}dy\right]\\
\nonumber &\times \int_{-1}^{1}dz\frac{1}{x^2+y^2+z^2}\, ,\\
\nonumber I_3=&-\frac{1}{16E_R}\left[\int_{-2}^{-1}dx+\int_{1}^{2}dx\right]\left[\int_{-2}^{-1}dy+\int_{1}^{2}dy\right]\\
\nonumber &\times \left[\int_{-2}^{-1}dz+\int_{1}^{2}dz\right]\frac{1}{x^2+y^2+z^2}\, .
\end{align}
Specifically, we have that $\bar{\chi}_{\mathbf{11};S=2}^{\mathbf{0}}=3I_1+3I_2+I_3$.  However, we note that this is
\begin{align}
\nonumber \bar{\chi}_{\mathbf{11};S=2}^{\mathbf{0}}=&-\frac{1}{16 E_R}\Big[\int_{-2}^{2}dx\int_{-2}^2dy\int_{-2}^{2}dz\frac{1}{x^2+y^2+z^2}\\
&-\int_{-1}^{1}dx\int_{-1}^1dy\int_{-1}^{1}dz\frac{1}{x^2+y^2+z^2}\Big]\\
&=\bar{\chi}_{\mathbf{11};S=1}^{\mathbf{0}}
\end{align}
provided that $\Lambda\ge 2$.  Following this construction for higher shells, we have the result that
\begin{align}
\sum_{\ell=1}^{S} \bar{\chi}_{\mathbf{11};\ell }^{\mathbf{0}}&=\min\left(S,\Lambda\right)\bar{\chi}_{\mathbf{11};S=1}^{\mathbf{0}}\, .
\end{align}
In Sec.~\ref{sec:MatrixElem}, we showed that the regularization parameter $\Lambda$ may be interpreted as a cutoff in Fourier components, see Eq.~\eqref{eq:hover}.  The present analysis demonstrates that $\Lambda$ can also be interpreted, albeit more loosely, as a cutoff in the contributions from higher bands.  As the lattice potential vanishes, higher bands become equivalent to plane waves of high energy.  Hence, the given lattice renormalization procedure produces the same results as the free-space renormalization procedure, Eq.~\eqref{eq:freerenorm}, where the cutoff $K^{\star}\sim \Lambda\pi /a$.  For the first term in the definition of $\chi^{\mathbf{K}}\left(E_{\mathbf{K}}\right)$ which includes the optical lattice, see Eq.~\eqref{eq:chidef}, $\Lambda$ is not a strict cutoff for the contributions of higher bands because the expansions Eq.~\eqref{eq:uFourier} do not contain only a single Fourier component.  This is why it is important to first take the limit $S\to\infty$ with $\Lambda$ fixed and then take $\Lambda\to \infty$.  However, the contributions to the integral from shells $S>\Lambda$ quickly approach their free-space values, causing the shell summation of $\chi^{\mathbf{K}}\left(E_{\mathbf{K}}\right)$ to converge, see Fig.~\ref{fig:ScalinginLambdaandS}.

\section{The Fermi resonance Hamiltonian}
\label{sec:FRHPRA:manybody}
We now proceed to step 2 of the scheme outlined in Fig.~\ref{fig:FRHPRA:schematic}, the partitioning into low and high energy spaces.  Typical operating temperatures of ultracold atomic gas experiments are below the band gap of the confining lattice potential, see Sec.~\ref{sec:EnergyScales}.  For moderate lattice strengths, this implies that fermions remain in the lowest open channel band when they are separated by a distance larger than the effective range $r_B\ll a$ to minimize their energy, and this defines the low-energy, long-wavelength sector of Hilbert space.  Only when fermions come together at very close distance comparable to the effective range are they able to populate excited bands of the open channel or transfer population to closed channel bands.  At low lattice fillings, the particular population of the high energy sector of Hilbert space is set by the two-body physics.  We find the relevant states from the high-energy sector of Hilbert space by solving the two-body problem projected into the high-energy subspace.  We call the eigenstates of the high-energy sector \emph{dressed molecules}.  The effective many-body description is then a resonance model between unpaired fermions in the lowest band and dressed molecules.  By the construction of the dressed molecules, this model reproduces the correct scattering and bound state properties at the two-particle level, and hence is expected to provide an accurate many-body description at low densities.

We refer to the resulting model as an effective model in the spirit of effective models such as the $t-J$ model~\cite{Auerbach_94} in that we have removed high-energy degrees of freedom and kept only the most relevant couplings for the low energy physics.  While excited bands are still present in the theory in the form of the dressed molecules they are restricted to occur only in specific combinations, just as quarks are required to be bound at low energies.  In solid state systems, often little is known about the microscopic Hamiltonian, and so the construction of an effective Hamiltonian may not follow a well-controlled prescription.  In contrast, the derivation of our effective model follows in a step-by-step manner from the microscopic physics.  Our partitioning into low and high energy sectors is accomplished using projectors $\hat{L}$ into the lowest open channel band and $\hat{D}=1-\hat{L}$ into all excited open channel bands as well as the closed channel molecules.  A similar approach was taken in Ref.~\cite{von_Stecher_Gurarie_11}.  An analysis parallel to Sec.~\ref{sec:FRHPRA:fewbody} gives a nonlinear eigenequation for the closed channel components of the high-energy sector as
\begin{align}
\label{eq:FRHPRA:NLDEE}&\left[E_{\mathbf{K}}-h\nu-{E}_{\mathbf{s}\mathbf{K}}^M\right]\Upsilon_{\mathbf{s}}^{\mathbf{K}}=\frac{g^2}{v}\sum_{\mathbf{t}}\tilde{\chi}_{\mathbf{st}}^{\mathbf{K}}\left(E_{\mathbf{K}}\right)\Upsilon_{\mathbf{t}}^{\mathbf{K}}\, ,\\
\label{eq:FRHPRA:Dchidef}&\tilde{\chi}_{\mathbf{st}}^{\mathbf{K}}\left(E_{\mathbf{K}}\right)\equiv\left[\sum_{\mathbf{nm}}'\int\frac{d\mathbf{q}}{v_{\mathrm{BZ}}}\frac{h_{\mathbf{s}\mathbf{K}}^{\mathbf{nm}}\left(\mathbf{q}\right) {h_{\mathbf{t}\mathbf{K}}^{\mathbf{nm}}}^{\star}\left(\mathbf{q}\right)}{E_{\mathbf{K}}-E_{\mathbf{nm}}^{\mathbf{K}}\left(\mathbf{q}\right)+i\eta}-\bar{\chi}_{\mathbf{st}}^{\mathbf{K}}\right]\, ,
\end{align}
where the prime on the summation denotes that the sum is taken over all $\left(\mathbf{n},\mathbf{m}\right)\ne \left(\mathbf{1},\mathbf{1}\right)$.  That is, the sum does not include the lowest open channel scattering band.  The renormalization matrix $\bar{\chi}^{\mathbf{K}}$ appearing in Eq.~\eqref{eq:FRHPRA:Dchidef} is the same as in Eq.~\eqref{eq:chidef}, and so the detuning and scattering length appearing parametrically in this projected model are the renormalized detuning and scattering length of the full two-body problem using all of the open channel bands.  In this way the projected system reproduces the correct, renormalized physics at the two-body level when recombined with the lowest scattering band of the open channel.

\subsection{Properties of the projected model}

\begin{figure}[thbp]
\centerline{\includegraphics[width=\columnwidth]{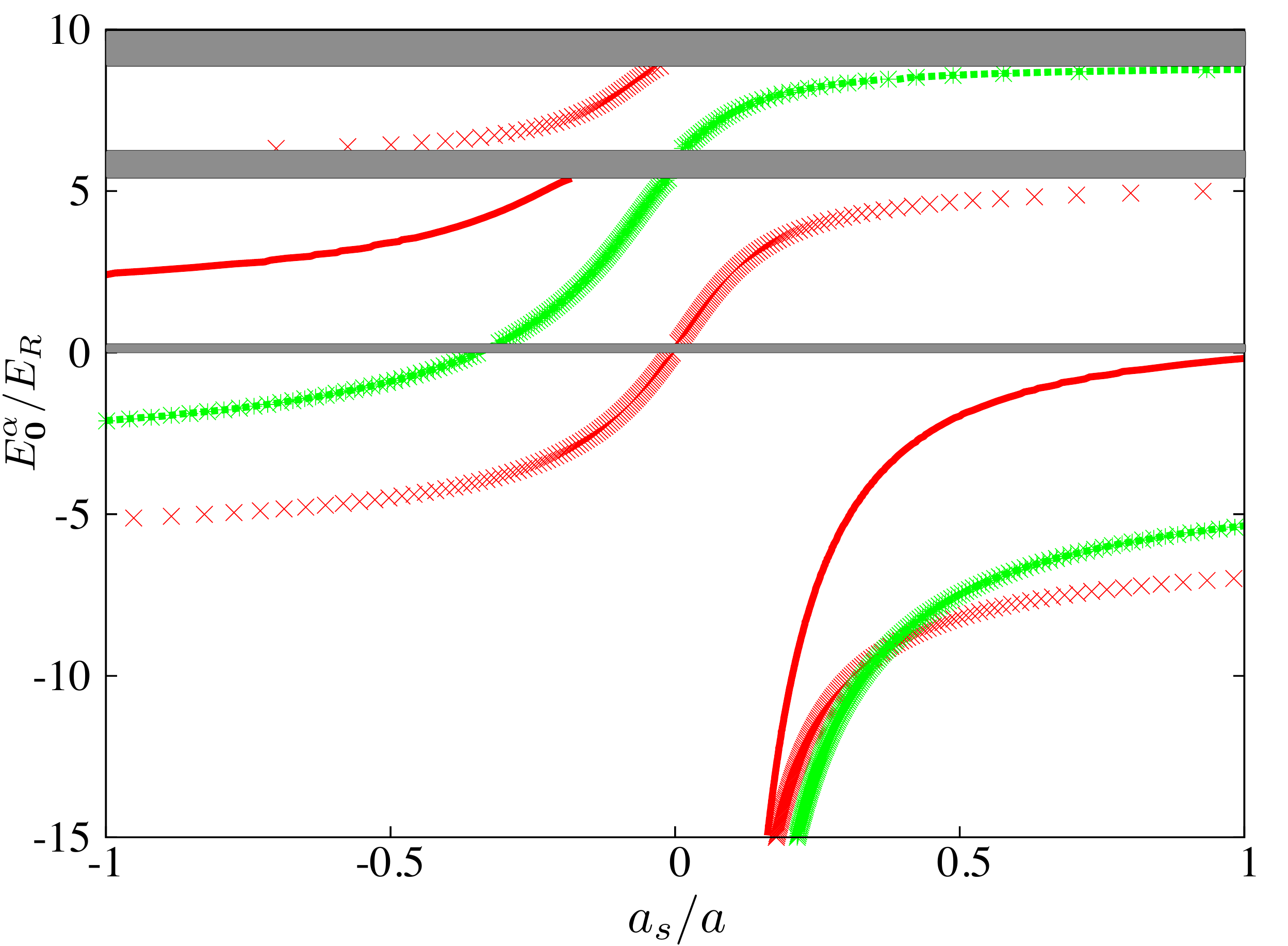}}
\caption{\label{fig:PrjCOmp} (Color online) \emph{Comparison of the zero-quasimomentum bound states of the two-body solution and the projected two-body solution.}  All results are for a 3D cubic optical lattice of strength $V=12E_R$, and the grey stripes denote scattering bands of the open channel.  The red crosses represent the two-particle bound states with completely even parity under inversion, and the green asterisks have parity $\left(1,1,-1\right)$ and cyclic permutations.  The solid red lines are the bound states of the projected model with completely even parity and the green dashed lines are the bound states of the projected model with parity $\left(1,1,-1\right)$ and cyclic permutations.  The odd parity solutions of the projected and full models very nearly coincide, but the resonance behavior in the completely even parity solution is drastically altered by the projection.}
\end{figure}

The bound state energies of the non-projected two-particle problem were displayed as a band structure in Ref.~\cite{wall_carr_12} where the Fermi resonance Hamiltonian was first identified.  In this section we examine the bound states of the projected two-particle problem for comparison.  For clarity of exposition, we will study the problem only at zero total quasimomentum.  The results of the comparison are shown in Fig.~\ref{fig:PrjCOmp}, which shows the lowest bound state energies of the two-particle problem both with and without projection as a function of the $s$-wave scattering length in the limit of an infinitely broad resonance, see Appendix \ref{sec:NLEEsoln}.  The red crosses and solid red lines represent solutions with completely even parity under inversion $\mathbf{p}=\left(1,1,1\right)$ of the non-projected and projected models, respectively.  The green crosses and dashed green lines represent solutions with parity $\mathbf{p}=\left(1,1,-1\right)$ of the non-projected and projected models, respectively.  The results for the odd parity bound states are almost identical between the non-projected and projected models, which is a statement that this bound state does not couple strongly to the lowest open channel scattering band.  In contrast, the lowest energy bound state with completely even parity displays drastic differences between the projected and non-projected models.  The resonance position of the projected model, which is the energy with respect to the bottom of the lowest open channel scattering continuum where the $s$-wave scattering length diverges, is shifted close to zero, indicating that this bound state causes a two-body resonance when energetically close to the threshold of the lowest band of the open channel.  This is the sense in which the dressed molecules form an effective closed channel for the two-body resonance near the lowest scattering band of the open channel.

As discussed in Ref.~\cite{wall_carr_12}, the asymptotic scaling of the matrix $\chi^{\mathbf{K}}$ with the regularization cutoff Eq.~\eqref{eq:FRHPRL:chiscaling} holds for the projected matrix $\tilde{\chi}^{\mathbf{K}}\left(E_{\mathbf{K}}\right)$, as well.  This follows from Eq.~\eqref{eq:hover}, which demonstrates that as $\Lambda$ becomes larger than the support of the Fourier expansions of the lowest open channel band Bloch functions the overlaps $h_{sK}^{11}\left(q\right)$ are no longer functions of $\Lambda$.  Hence, the scaling with $\Lambda$ is dominated by the contributions of higher bands.  The same argument extends to a projected matrix $\tilde{\chi}^{\mathbf{K}}$ in which any finite number of open channel bands have been projected out.  These results reinforce the interpretation put forth in Sec.~\ref{sec:properreg} that the effect of the regularization is a cutoff in the contributions from higher bands, and so the results for large $\Lambda$ should be insensitive to the behavior of low-energy bands.

We conclude this section by noting that the method of projection could be applied in principle to some other scattering continuum $\left(\mathbf{n},\mathbf{m}\right)$, in which case the bound state which couples most strongly to that continuum would have its resonance shifted such that the scattering length diverges at approximately $\min\left\{E_{\mathbf{nm}}^{\mathbf{K}}\left(\mathbf{q}\right), \mathbf{q}\in\mathrm{BZ},\mathbf{K}\in\mathrm{BZ}\right\}$.  This allows for the derivation of effective Hamiltonians for resonances in excited bands.  Additionally, one can consider the first $n$ scattering bands of the open channel to be the low energy sector of Hilbert space and project these out of the two-particle theory.  The dressed molecules of the projected high-energy sector of Hilbert space will define an effective closed channel which reproduces the scattering properties of the lowest $n$ open channel bands properly.  In this way, the construction can be extended to higher relative scattering energy where scattering states in higher bands can play a role.  In the remainder of this document we focus only on the case of resonance with the first scattering continuum.

\subsection{Wannier functions and Kohn construction}
\label{sec:Wannier}

We now move on to item 3 of the scheme outlined in Fig.~\ref{fig:FRHPRA:schematic}, the solution of the high-energy sector of Hilbert space.  The details of the numerical solution are discussed in Appendices~\ref{sec:GMA}-\ref{sec:NLEEsoln}.  In this section, we focus only on the structure of the solutions.  The two-particle solutions of the projected nonlinear eigenvalue problem Eq.~\eqref{eq:FRHPRA:NLDEE} may be written as
\begin{align}
\label{eq:Dtpstate}&\tilde{\Psi}_{\mathbf{K}\alpha}(\mathbf{x},\mathbf{y})=\frac{1}{\tilde{\mathcal{N}}_{\mathbf{K}\alpha}}\Big[\sum_{\mathbf{s}}\Upsilon^{\mathbf{K}}_{\mathbf{s}\alpha}\phi^{\left(b\right)}_{\mathbf{sK}}(\mathbf{x})\kappa(\mathbf{x}-\mathbf{y})\\
\nonumber&+\tilde{g}\sum_{\mathbf{nms};\mathbf{q}}'\frac{\Upsilon_{\mathbf{s}\alpha}^{\mathbf{K}}h_{\mathbf{sK}}^{\mathbf{nm}}(\mathbf{q})\phi^{\left(f\right)}_{\mathbf{nq}}(\mathbf{x})\phi^{\left(f\right)}_{\mathbf{m},\mathbf{K}-\mathbf{q}}(\mathbf{y})}{E_{\mathbf{K}}^{\alpha}-E_{\mathbf{nm}}^{\mathbf{K}}(\mathbf{q})}\Big]\, ,
\end{align}
where the prime on the summation indicates the exclusion of the lowest two-particle band of the open channel, compare Eq.~\eqref{eq:tpstate}, and the reduced Feshbach coupling $\tilde{g}$ was defined in Eq.~\eqref{eq:natunits1}.  Here, the normalization factor is
\begin{align}
\tilde{\mathcal{N}}_{\mathbf{K}\alpha}^2&=1-\tilde{g}^2\boldsymbol{\Upsilon}^{\mathbf{K}}_{\alpha}\cdot\frac{\partial \tilde{\chi}^{\mathbf{K}}(E_{\mathbf{K}}^{\alpha}/E_R)}{\partial E_{\mathbf{K}}}\cdot\boldsymbol{\Upsilon}^{\mathbf{K}}_{\alpha}\, .
\end{align}
  We define two-particle Wannier functions centered around some lattice site $i$ as the quasimomentum Fourier transforms of the two-particle solutions, 
\begin{align}
\label{eq:TPWannier}\mathcal{W}_{i\alpha}\left(\mathbf{x},\mathbf{y}\right)&=\frac{1}{\sqrt{N^3}}\sum_{\mathbf{K}\in\mathrm{BZ}}e^{-i\mathbf{K}\cdot\mathbf{R}_i}\tilde{\Psi}_{\mathbf{K}\alpha}(\mathbf{x},\mathbf{y})\, .
\end{align}
As is known from the seminal study of Kohn in 1D~\cite{Kohn_59}, the choice of phases for the Bloch functions affects the localization properties of the Wannier functions around the site $i$.  In 1D, choosing the Bloch functions to be smooth functions of the quasimomentum leads to exponential localization of the Wannier functions.  In the present case, the phases are chosen according to the construction of Sec.~\ref{sec:spandsymm} in which the eigenvectors \{$\boldsymbol{\Upsilon}^{\mathbf{K}}_{\alpha}$\} are smooth functions of the total quasimomentum except possibly at high-symmetry points of the BZ and phases under inversion are specified by the parity of the eigenstate.  Accounting for transformation under inversion is important to ensure that the parameters in the resulting many-body model have the same symmetries as the few-body physics.  The parity of the eigenstate is unambiguously defined at extremal points of the BZ in which all components of the total quasimomentum are either the zone center $K=0$ or the zone boundary $K=-\pi/a$, and then extended into the general BZ by the smoothness requirements on the eigenvectors $\Upsilon^{\mathbf{K}}_{\mathbf{s}\alpha}$.  We will refer to this choice of phases as the \emph{Kohn construction}.  More localized functions may be possible by replacing the quasimomentum Fourier transform in Eq.~\eqref{eq:TPWannier} with a more general unitary transformation which is consistent with inversion symmetry; we do not consider this possibility further here.

\subsection{Statement of the Hamiltonian and evaluation of parameters}

With the dressed molecule Wannier functions Eq.~\eqref{eq:TPWannier}, we can state the \emph{Fermi resonance Hamiltonian} (FRH) as~\cite{wall_carr_12}
\begin{align}
 \nonumber&\hat{H}_{\mathrm{eff}}=-t_f\sum_{\sigma\in\left\{\uparrow,\downarrow\right\}}\sum_{\langle i,j\rangle}\hat{a}_{i\sigma}^{\dagger}\hat{a}_{j\sigma}+E_0\sum_{\sigma\in\left\{\uparrow,\downarrow\right\}}\sum_i \hat{n}_{i\sigma}^{\left(f\right)}\\
 &\nonumber-\sum_{\alpha\in\mathcal{M}}\sum_{i,j}t^{\alpha}_{i,j}\hat{d}_{i\alpha}^{\dagger}\hat{d}_{j\alpha}+\sum_{\alpha\in\mathcal{M}}\bar{\nu}_{\alpha}\sum_i\hat{n}_{i\alpha}^{\left(b\right)}\\
\label{eq:FRHPRA:effHami}&+\sum_{\alpha\in\mathcal{M}}\sum_{ijk}g_{i-k,k-j}^{\alpha}\left[\hat{d}_{i,\alpha}^{\dagger}\hat{a}_{j\uparrow}\hat{a}_{k\downarrow}+\mathrm{h.c.}\right]\, .
\end{align}
Here, $\hat{a}_{i\sigma}$ is a fermionic operator which destroys a particle with spin $\sigma$ in the Wannier state $w_{i\sigma}\left(\mathbf{r}\right)$ constructed from the lowest band Bloch functions of the open channel, $\hat{n}_{i\sigma}^{\left(f\right)}=\hat{a}_{i\sigma}^{\dagger}\hat{a}_{i\sigma}$ is the number operator for particles with spin $\sigma$ in the lowest band of the open channel, $\hat{d}_{i\alpha}$ is a bosonic operator which destroys a dressed molecule in Wannier state $\mathcal{W}_{i\alpha}\left(\mathbf{x},\mathbf{y}\right)$, and $\hat{n}_{i\alpha}^{\left(b\right)}$ is the number operator for dressed molecules in state $\alpha$ on site $i$.  The summations over $\alpha\in\mathcal{M}$ indicate that the particular set of dressed molecules $\mathcal{M}$ which are relevant to describing the resonance properly depend on the open channel band being projected, symmetry considerations, and the energetic level of approximation used to truncate the terms appearing in Eq.~\eqref{eq:FRHPRA:effHami}.  For the case we focus on here in which the lowest open channel band is projected out, the lowest energy dressed molecule which has completely even parity under inversion is the most relevant one, as all others either have vanishing on-site effective pairing by symmetry or are far detuned from resonance~\cite{wall_carr_12}.

We now turn to the last part of the program outlined in Fig.~\ref{fig:FRHPRA:schematic} in which the low-energy and high-energy sectors of Hilbert space are re-coupled at the many-body level.  This re-coupling is performed by taking the expectation of the Hamiltonian within the basis of the open channel fermions in the lowest band and the dressed molecules, and provides us with the Hubbard parameters appearing in the FRH, Eq.~\eqref{eq:FRHPRA:effHami}.  The first term in Eq.~\eqref{eq:FRHPRA:effHami} represents tunneling of fermions with spin $\sigma$ in the lowest open channel band between neighboring sites $i$ and $j$.  The calculation of the tunneling using the Wannier functions for the lowest band fermions is well known~\cite{jaksch_bruder_98}.  The next term is the energy offset of a fermion in the lowest band with respect to the lowest two-particle scattering continuum, where $E_0=\int {d\mathbf{q}}\,E_{\mathbf{1},\mathbf{q}}/v_{\mathrm{BZ}}$.  The remaining three terms represent processes involving dressed molecules.  The first represents tunneling of the dressed molecular center of mass between two lattice sites $i$ and $j$.  The two sites $i$ and $j$ need not be nearest neighbors due to the nontrivial dispersion relation of the dressed molecules.  The next term represents the effective detuning of a dressed molecular Wannier function away from resonance with the lowest two-particle scattering band of the open channel.  The final term is the effective pairing of two open channel fermions into a dressed molecule at the center of mass.

The dressed molecular tunneling Hubbard parameter is evaluated as
\begin{align}
\nonumber t_{i,j}^{\alpha}&=-\int d\mathbf{x}d\mathbf{y}\mathcal{W}^{\star}_{i\alpha}\left(\mathbf{x},\mathbf{y}\right)\hat{{H}}\,\mathcal{W}_{j\alpha}\left(\mathbf{x},\mathbf{y}\right)\,  ,\\
&=-\int \frac{d\mathbf{K}}{v_{\mathrm{BZ}}}e^{i\mathbf{K}\cdot\left(\mathbf{R}_i-\mathbf{R}_j\right)}E_{\mathbf{K}}^{\alpha}\, ,
\end{align}
where $\hat{{H}}$ is the two-particle Hamiltonian.  The fact that the tunneling does not change the internal state $\alpha$ of the dressed molecules is a consequence of the fact that the dressed molecules diagonalize the high-energy sector of the Hamiltonian.  The remarkable fact that the tunneling of these very strongly correlated objects is a simple quasimomentum Fourier transform of their dispersion is due to the invariance of the entire system under translation.

The dressed molecular tunneling has several novel features compared to single-particle tunneling.  The interaction destroys the separability of the band structure along principal axes and so tunneling is allowed along directions which are not the principal axes of the lattice.  This is in stark contrast to single-particle tunneling, which always occurs along the principal axes.  Additionally, for traps where the center of mass and relative motion separate, diagonal tunneling will always vanish.  Thus, diagonal tunneling is a unique feature of the FRH due to non-separability in the dressed closed channel.  While we have never found the diagonal tunneling to be greater than or equal to the nearest neighbor tunneling, it is often as large or larger than the second neighbor tunneling and in some cases can be of the same order of magnitude as the nearest neighbor tunneling.  Second, the dispersion relation of the dressed molecules is highly nontrivial due to the fact that dressed molecules consist of two strongly interacting particles, and so the sign of the tunneling depends on the parity of the eigenstate and the direction and magnitude of the tunneling in a nontrivial way.  Quantitative values for the dressed molecular tunneling, including diagonal tunneling, are provided in Ref.~\cite{wall_carr_12}.

The effective detunings of the dressed molecules are evaluated as
\begin{align}
\nonumber \bar{\nu}_{\alpha}&=\int d\mathbf{x}d\mathbf{y}\mathcal{W}^{\star}_{i,\alpha}\left(\mathbf{x},\mathbf{y}\right)\hat{{H}}\,\mathcal{W}_{i,\alpha}\left(\mathbf{x},\mathbf{y}\right)\, ,\\
&=\int \frac{d\mathbf{K}}{v_{\mathrm{BZ}}}E_{\mathbf{K}}^{\alpha}\, ,
\end{align}
which is the average of the interacting band structure in the BZ.  The magnitudes of these detunings, compared with the pairing strengths to be discussed, determine which of the dressed molecules are nearby the lowest band in energy and so must be included dynamically in the FRH for a given $\tilde{a}_s$.  The Hubbard parameters for the pairing are evaluated as
\begin{align}
&g_{i-k,k-j}^{\alpha}=\int d\mathbf{x}d\mathbf{y} \mathcal{W}^{\star}_{i,\alpha}\left(\mathbf{x},\mathbf{y}\right)\hat{H}w_j\left(\mathbf{x}\right)w_k\left(\mathbf{x}\right)\, .
\end{align}
The open channel fermions couple to the dressed molecules only through Feshbach coupling to their closed channel components according to the two-channel model, see Eqs.~(\ref{eq:SE1}-\ref{eq:SE2}).  Thus,
\begin{align}
\nonumber \frac{g_{i-k,k-j}^{\alpha}}{E_R}=&g\int d\mathbf{x}\frac{1}{\sqrt{N^3}}\sum_{\mathbf{K}\in\mathrm{BZ}}\frac{e^{i\mathbf{K}\cdot\mathbf{R}_i}}{E_R\tilde{\mathcal{N}}_{\mathbf{K}\alpha}}\\
&\times \sum_{\mathbf{s}}\Upsilon_{\mathbf{s}\alpha}^{\mathbf{K}}\phi^{\left(b\right)\star}_{\mathbf{sK}}\left(\mathbf{x}\right)w_j\left(\mathbf{x}\right)w_k\left(\mathbf{x}\right)\, ,\\
\nonumber =& \tilde{g}\int \frac{d\mathbf{K}}{v_{\mathrm{BZ}}} \frac{e^{i\mathbf{K}\cdot \boldsymbol{\Delta}_{ik}}}{\tilde{\mathcal{N}}_{\mathbf{K}\alpha}}\\
\label{eq:effpairing}&\times \int \frac{d\mathbf{q}}{v_{\mathrm{BZ}}} e^{i\mathbf{q}\cdot \boldsymbol{\Delta}_{kj}}\sum_{\mathbf{s}}\Upsilon_{\mathbf{s}\alpha}^{\mathbf{K}} h_{\mathbf{sK}}^{\mathbf{11}}\left(\mathbf{q}\right)\, ,
\end{align}
where we have defined $\boldsymbol{\Delta}_{ij}\equiv\mathbf{R}_i-\mathbf{R}_j$ and used the definition Eq.~\eqref{eq:hoverlaps} in the last line.  Hence, the effective coupling constant at total quasimomentum $\mathbf{K}$, $g_{\alpha\mathbf{K}}$, is
\begin{align}
\frac{g_{\alpha\mathbf{K}}}{E_R}&=\tilde{g}(1-\tilde{g}^2\boldsymbol{\Upsilon}^{\mathbf{K}}_{\alpha}\cdot\frac{\partial \tilde{\chi}^{\mathbf{K}}(E_{\mathbf{K}}^{\alpha}/E_R)}{\partial E_{\mathbf{K}}}\cdot\boldsymbol{\Upsilon}^{\mathbf{K}}_{\alpha})^{-1/2}\, .
\end{align}
In the limit of an infinitely broad resonance where $\tilde{g}\to \infty$, we have that
\begin{align}
\lim_{\tilde{g}\to \infty}\frac{g_{\alpha\mathbf{K}}}{E_R}&=(-\boldsymbol{\Upsilon}^{\mathbf{K}}_{\alpha}\cdot\frac{\partial \tilde{\chi}^{\mathbf{K}}(E_{\mathbf{K}}^{\alpha}/E_R)}{\partial E_{\mathbf{K}}}\cdot\boldsymbol{\Upsilon}^{\mathbf{K}}_{\alpha})^{-1/2}\, ,
\end{align}
which depends on the divergent quantity $\tilde{g}$ only parametrically through the derivative of the matrix $\tilde{\chi}^{\mathbf{K}}$.  Physically, as the resonance becomes very broad, the closed channel population of the eigenstate vanishes, see Eq.~\eqref{eq:Dtpstate}.  However, the interchannel coupling diverges in this same limit, and the infinite coupling to an infinitesimal closed channel population produces a finite limit.  In Appendix~\ref{appendix:chider} it is proved that the matrix $-{\partial \tilde{\chi}^{\mathbf{K}}(E_{\mathbf{K}}^{\alpha}/E_R)}/{\partial E_{\mathbf{K}}}$ is positive semidefinite and so the pairing is real and the transformation to the FRH has the effect of narrowing the resonance irrespective of the value of the bare pairing strength $g$.  The behavior of the effective pairing Eq.~\eqref{eq:effpairing} in typical cases is given in Ref.~\cite{wall_carr_12}.

We note that, while the FRH is still in general a multi-band Hamiltonian with the eigenstate index $\alpha$ as the band index, in typical cases we need only a single dressed molecular band as compared to the large number of bands required for an accurate calculation without dressing the closed channel.  For example, for the case of a resonance in the lowest band of the open channel, the lowest energy dressed molecule with completely even parity is the most relevant to describing the resonance behavior.  The dressed molecules thus form an ideal dressed closed channel, and the FRH is the two-channel lattice model between fermions in the lowest band and these dressed molecules.  Dressed molecules which are off-resonant can be adiabatically eliminated from the theory, and their effects can be included as a background scattering term using e.g.~second order perturbation theory.  The possibility of three- and higher-body effects are not captured by the two-channel model, and so such terms do not appear in the FRH as written.  However, this is a shortcoming of the continuum modeling, step 1 from Fig.~\ref{fig:FRHPRA:schematic}, and can be overcome by performing the lattice projection of a more complex continuum scattering model.

\section{Conclusions}
\label{sec:FRHPRA:conclusion}

We have presented a complete two-particle theory of two-component fermions in a three-dimensional lattice potential interacting through a Feshbach resonance within the confines of the two-channel model.  The two-particle lattice bound states are obtained from an eigenvalue equation which is nonlinear in the energy eigenvalue.  Specializing to the case of a simple cubic lattice, we showed that the matrix elements of the interchannel coupling in the Bloch basis have a simple representation in terms of the Fourier components of the Bloch basis.  This Fourier expansion may be found efficiently numerically by solving a real symmetric eigenvalue problem.  Analyzing the interchannel overlaps, we demonstrated that the effect of an interchannel regularization function which has the same symmetries as the lattice may be interpreted in terms of a Fourier cutoff in a shifted momentum space which has the symmetries of the momentum-space BZ.  This result unifies the understanding of the renormalization of two-channel theories in the lattice and in free space, as renormalization in the latter amounts to imposing a spherically symmetric cutoff in momentum space.  Using the simplified expressions for the interchannel overlaps, we computed the lattice renormalization at zero total quasimomentum, and demonstrated that the renormalization may also be interpreted as a cutoff in the contributions from higher Bloch bands.  We presented a scaling relation for the effective closed channel $T$-matrix which allows results for finite regularization cutoff to be extrapolated to infinite cutoff.  By arranging limits appropriately, this scaling procedure produces results for the bound states which properly accounts for the effects of all higher bands.

The two-particle theory was then applied to a two-channel model in which the lowest open channel scattering band was projected out.  We called these bound states \emph{dressed molecules}.  When recombined with the lowest band of the open channel, the dressed molecules accurately reproduce the resonance physics at the two-body level, including proper renormalization and the effects of higher bands.  We analyzed the dressed molecular dispersion as a function of $s$-wave scattering length at zero total quasimomentum, and showed that their interpretation as an effective closed channel for the two-body resonance is valid.  We then defined localized Wannier functions from these two-body states and gave a phase prescription such that the Wannier functions transform with the symmetries of the lattice.  Taking matrix elements of the many-body Hamiltonian in the basis of the dressed molecular Wannier functions, we defined the \emph{Fermi resonance Hamiltonian} (FRH).    The FRH has the interpretation of a many-body resonance model between unpaired fermions in the lowest band and dressed molecules.  By symmetry and energetic considerations, only a single dressed molecule is relevant in typical cases, and so the FRH often reduces to an effective two-channel model.  The use of the full lattice solution to determine the dressed molecular Wannier functions not only increases the quantitative predictive power of the model versus models which use separable approximations to the lattice, but also leads to interesting features of the bound states such as tunneling along non-principal axes.

The results in this work can be applied and extended to a variety of scenarios.  The problem of pairing of unequal-mass fermions is relevant to the formation of ultracold molecules via magneto-association across a Feshbach resonance~\cite{lercherAD2011}.  The interesting case of quasi-low dimensional systems with strong $s$-wave interactions can be studied by altering the lattice depths along the principal axes in the lattice potential, see Eq.~\eqref{eq:Lattdef}.  In a similar vein, our results can be extended to arbitrary lattice geometries.  In particular, our derivation of the nonlinear eigenvalue equation for bound states and our exposition of the renormalization in terms of a momentum cutoff  which has the same shape as the reciprocal-space BZ apply to any lattice geometry.  Finally, more general interchannel potentials can be accommodated, for example to study pairing to higher angular momentum states.

We acknowledge useful discussions with John Bohn, Hans Peter B\"{u}chler, Charles Clark, Daniel Schirmer, David Wood, and Zhigang Wu.  We also thank Hans Peter B\"{u}chler for sharing computer code with us during early stages of this work.  This work was supported by the Alexander von Humboldt Foundation, AFOSR grant number FA9550-11-1-0224,  the Heidelberg Center for Quantum Dynamics, and the National Science Foundation under Grants PHY-1207881, PHY-1067973, and PHY-0903457.  We also acknowledge the Golden Energy Computing Organization at the Colorado School of Mines for the use of resources acquired with financial assistance from the National Science Foundation and the National Renewable Energy Laboratories.

\appendix

\section{Properties of the derivative of the effective closed channel $T$-matrix}
\label{appendix:chider}
Here we wish to prove that $-{\partial \chi^{\mathbf{K}}}/{\partial E}$ is positive definite outside of open channel scattering bands.  As we consider ourselves to be outside of scattering bands, we drop the infinitesimal $\eta$ in Eq.~\eqref{eq:chidef}.  We note that we may write $-{\partial \chi^{\mathbf{K}}}/{\partial E}$ as the Gram matrix
\begin{align}
-\frac{\partial \chi^{\mathbf{K}}}{\partial E}&=\langle \mathbf{f}_{\mathbf{sK}},\mathbf{f}_{\mathbf{tK}}\rangle
\end{align}
by defining the vectors
\begin{align}
{f}_{\mathbf{sK}}^{\mathbf{nmq}}&\equiv\frac{h_{\mathbf{s}\mathbf{K}}^{\mathbf{nm}}\left(\mathbf{q}\right)}{E_{\mathbf{K}}-E_{\mathbf{nm}}^{\mathbf{K}}\left(\mathbf{q}\right)}\, .
\end{align}
This proves that $-{\partial \chi^{\mathbf{K}}}/{\partial E}$ is positive semidefinite.  Positive definiteness follows from the fact that the set $\left\{\mathbf{f}_{\mathbf{sK}}\right\}$ are linearly independent.  We prove this by first noting that the parameterization Eq.~\eqref{eq:openchannel} of functions which transform \emph{$K$-periodically} as $\psi\left(\mathbf{x}+\mathbf{a},\mathbf{y}+\mathbf{a}\right)=e^{i\mathbf{K}\cdot\mathbf{a}}\psi\left(\mathbf{x},\mathbf{y}\right)$ for any Bravais lattice vector $\mathbf{a}$ implies that
\begin{align}
\nonumber&\sum_{\mathbf{nmq}}\phi^{\star}_{\mathbf{nq}}\left(\mathbf{x}\right)\phi_{\mathbf{nq}}\left(\mathbf{x}'\right)\phi^{\star}_{\mathbf{m},\mathbf{K}-\mathbf{q}}\left(\mathbf{y}\right)\phi_{\mathbf{m},\mathbf{K}-\mathbf{q}}\left(\mathbf{y}'\right)\\
&\;\;\;\;\;\;\;\;=\delta\left(\mathbf{x}-\mathbf{x}'\right)\delta\left(\mathbf{y}-\mathbf{y}'\right)
\end{align}
within the space of $K$-periodic functions.  That is to say, products of Bloch functions with fixed total quasimomentum form a complete basis for the space of functions which transform with this total quasimomentum under translations.  This implies that
\begin{align}
\label{eq:completenessrelation}\lim_{\Lambda\to\infty} \sum_{\mathbf{nmq}}h_{\mathbf{sK}}^{\mathbf{nm}}\left(\mathbf{q}\right)h_{\mathbf{tK}}^{\mathbf{nm}\star}\left(\mathbf{q}\right)&=\delta_{\mathbf{s},\mathbf{t}}\, .
\end{align}
The elements of $h_{\mathbf{sK}}^{\mathbf{nm}}\left(\mathbf{q}\right)$ with $\mathbf{s}$ and $\mathbf{K}$ fixed, arranged as a vector, $\mathbf{h}_{\mathbf{sK}}$, thus form an orthogonal set, and so the determinant of the matrix with these vectors as rows is nonzero.  If we now scale the $\mathbf{nmq}$ element of each vector $\mathbf{h}_{\mathbf{sK}}$ by the same factor $\left(E_{\mathbf{K}}-E_{\mathbf{nm}}^{\mathbf{K}}\left(\mathbf{q}\right)\right)^{-1}$, this determinant is multiplied by the product of all of these factors.  Hence, the determinant is nonzero as long as each of the factors is nonzero.  The factors are nonzero so long as the energy $E_{\mathbf{K}}$ is finite and does not lie in one of the scattering bands of the open channel.  The scaled vectors are the set $\left\{\mathbf{f}_{\mathbf{sK}}\right\}$, and the non-vanishing determinant implies that this set is linearly independent, completing the proof.

For the projected model, we instead deal with $-{\partial \tilde{\chi}^{\mathbf{K}}}/{\partial E}$ and the relation Eq.~\eqref{eq:completenessrelation} no longer applies, as we have excluded the lowest band of the open channel.  Hence, our results show only that $-{\partial \tilde{\chi}^{\mathbf{K}}}/{\partial E}$ is positive semidefinite.  In all numerical cases we have considered, the scalar product $-\boldsymbol{\Upsilon}^{\mathbf{K}}_{\alpha}\cdot \frac{\partial \tilde{\chi}^{\mathbf{K}}}{\partial E} \cdot \boldsymbol{\Upsilon}^{\mathbf{K}}_{\alpha}$ is nonzero.

\section{Construction and scaling of the effective closed channel $T$-matrix}
\label{sec:GMA}
The most numerically demanding part of the two-particle solution is the construction of the matrix $\chi^{\mathbf{K}}\left(E_{\mathbf{K}}\right)$ defined in Eq.~\eqref{eq:chidef}.  While the overlaps $h_{\mathbf{sK}}^{\mathbf{nm}}\left(\mathbf{q}\right)$ may be written as products of 1D functions for the simple cubic lattice, see Sec.~\ref{sec:MatrixElem}, the construction of $\chi^{\mathbf{K}}\left(E_{\mathbf{K}}\right)$ does not factorize due to a nontrivial denominator.  In this section, we present an overview of efficient numerical methods for computing the values of $\chi^{\mathbf{K}}\left(E_{\mathbf{K}}\right)$ and for performing the scaling Eq.~\eqref{eq:FRHPRL:chiscaling} necessary for a properly renormalized theory.

We begin with a discussion of the integration over the BZ in Eq.~\eqref{eq:chidef}.  Because little is known generally about the behavior of the integrand for arbitrary $\mathbf{K}$ and band indices, it is best to look for an adaptive approach which refines the integration grid until a desired numerical tolerance is met.  For the cubic BZ that we consider, a natural choice is the Genz-Malik algorithm~\cite{Genz_Malik_80}, an adaptive integration method over hyper-rectangular regions in 3D.  We define a generator $F\left[I,\mathbf{v}\right]$ which takes as input a function $I\left(\mathbf{v}\right): \mathbb{R}^3\to \mathbb{R}$ and a vector $\mathbf{v}\in\mathbb{R}^3$ as the sum of function evaluations of $I$ at all unique permutations of the elements of $\mathbf{v}$ including sign changes.  For example,
\begin{align}
&F\left[I,\left(0,0,0\right)\right]=I\left(\left(0,0,0\right)\right)\, ,\\
\nonumber &F\left[I,\left(1,0,0\right)\right]=I\left(\left(1,0,0\right)\right)+I\left(\left(0,1,0\right)\right)+I\left(\left(0,0,1\right)\right)\\
&+I\left(\left(-1,0,0\right)\right)+I\left(\left(0,-1,0\right)\right)+I\left(\left(0,0,-1\right)\right)\, .
\end{align}
The Genz-Malik algorithm uses the ansatz
\begin{align}
\nonumber &\int_{-1}^{1}dx\int_{-1}^{1}dy\int_{-1}^{1}dz I\left(x,y,z\right)\approx R_7\left(I,{w},{\lambda}\right)\, ,\\
\label{eq:GMR7}&R_7\left(I,{w},{\lambda}\right)\equiv w_1F\left[I,\left(0,0,0\right)\right]\\
\nonumber &+w_2F\left[I,\left(\lambda_2,0,0\right)\right]+w_3F\left[I,\left(\lambda_3,0,0\right)\right]\\
&+w_4F\left[I,\left(\lambda_4,\lambda_4,0\right)\right]+w_5F\left[I,\left(\lambda_5,\lambda_5,\lambda_5\right)\right]\, ,
\end{align}
and then chooses the parameters $\left\{w_i\right\}$ and $\left\{\lambda_i\right\}$ such that $R_7$ is a rule of degree seven, meaning that it exactly integrates all monomials of degree at most seven~\cite{Stroud_71}.  This is in analogy to the more familiar Gauss-Legendre quadrature in 1D, where a set of integration points and weights are chosen such that the quadrature rule exactly integrates all polynomials of a given degree~\cite{press1993}.  An appropriate choice of parameters is~\cite{Genz_Malik_80}
\begin{align}
\nonumber w_1&=-\frac{87488}{19683}\, ,\;\;w_2=\frac{7840}{6561}\, ,\;\;w_3=\frac{4960}{19683}\, ,\\
w_4&=\frac{1600}{19683}\, ,\;\;w_5=\frac{6859}{19683}\, ,\\
\lambda_2^2&=\frac{9}{70}\, ,\;\;\lambda_3^2=\lambda_4^2=\frac{9}{10}\, ,\;\;\lambda_5^2=\frac{9}{19}\, .
\end{align}
The domain of integration can be deformed to any hyper-rectangular region by appropriate scaling of the integrand.  Additionally, as the integration points do not include the boundaries, the domain of integration can be constructed so as to avoid troublesome function evaluations for singular integrands.  To estimate the error of the rule $R_7\left(I,w,\lambda\right)$ a rule of degree five, $R_5\left(I,w',\lambda\right)$, is constructed using only function evaluations at the points $\left\{\lambda_i\right\}$ as
\begin{align}
\nonumber R_5\left(I,w',\lambda\right)&\equiv w_1'F\left[I,\left(0,0,0\right)\right]+w_2'F\left[I,\left(\lambda_2,0,0\right)\right]\\
\label{eq:GMR5}&+w_3'F\left[I,\left(\lambda_3,0,0\right)\right]+w_4'F\left[I,\left(\lambda_4,\lambda_4,0\right)\right]\, ,
\end{align}
with
\begin{align}
w_1'&=-\frac{4456}{243}\, ,\;\;w_2'=\frac{980}{243}\, ,\;\;w_3'=-\frac{140}{729}\, ,\;\;w_4'=\frac{200}{729}\, .
\end{align}

The algorithm begins by dividing the integration region into hyper-rectangular regions, and then using Eqs.~\eqref{eq:GMR7} and \eqref{eq:GMR5} to provide degree 7 and 5 estimates of the integral in each region, $R_7$ and $R_5$.  The error in each region is estimated by the absolute difference $\left|R_7-R_5\right|$, and the region with greatest error is identified.  We now wish to subdivide the region of greatest error into two hyper-rectangular regions by dividing the region in half along a Cartesian dimension.  The dimension to be subdivided is the one with the greatest fourth difference of the integrand, estimated as, e.g.,
\begin{align}
D_x&^{\left(4\right)}I\left(\mathbf{v}\right)\Big|_{\mathbf{v}=\left(0,0,0\right)}\\
\nonumber \approx& I\left(\left(\lambda_2,0,0\right)\right)+I\left(\left(-\lambda_2,0,0\right)\right)-2I\left(\left(0,0,0\right)\right)\\
\nonumber &-\frac{\lambda_2^2}{\lambda_4^2}\left[ I\left(\left(\lambda_4,0,0\right)\right)+I\left(\left(-\lambda_4,0,0\right)\right)-2I\left(\left(0,0,0\right)\right)\right]\, .
\end{align}
Here, we have placed the center of the region to be subdivided at $\mathbf{v}=(0,0,0)$ for simplicity.  We note that the fourth derivative estimation uses only points required in evaluating $R_7$ and $R_5$.  Estimates for the integral on the new subregions are provided by applying Eqs.~\eqref{eq:GMR7} and \eqref{eq:GMR5}, and the process of finding the region of largest error, subdividing according to the greatest fourth difference, and computing integrals on the new subregions is continued until the total error is bounded by a user-defined tolerance.  A nice feature of the Genz-Malik algorithm is that it parallelizes well over subregions~\cite{Berntsen_Espelid_91}.  Additionally, as discussed in Sec.~\ref{sec:MatrixElem}, the matrix elements $h_{\mathbf{sK}}^{\mathbf{nm}}\left(\mathbf{q}\right)$ separate along principal axes and so the computation of these functions may be cached along each Cartesian dimension to avoid overhead in re-computation.  Similar algorithms can be devised for non-rectangular regions~\cite{Genz91anadaptive} to enable the study of the FRH on other lattice structures.

As discussed in Sec.~\ref{sec:properreg}, proper computation of $\chi^{\mathbf{K}}\left(E_{\mathbf{K}}\right)$ first fixes $\Lambda$ and converges a summation as the number of open channel bands $S\to\infty$ before letting $\Lambda\to\infty$ according to the scaling relationship Eq.~\eqref{eq:FRHPRL:chiscaling}.  In practice we use shell summation as defined in Sec.~\ref{sec:properreg} to converge $\chi^{\mathbf{K}}\left(E_{\mathbf{K}}\right)$ as $S\to\infty$ to a desired relative precision $\epsilon_{\mathrm{rel}}$.  The diagonal elements $\chi_{\mathbf{ss}}^{\mathbf{K}}\left(E_{\mathbf{K}}\right)$ are typically larger than the off-diagonal elements, and so it may be desirable to put an absolute tolerance $\epsilon_{\mathrm{abs}}$ on the off-diagonal elements which is comparable to $\epsilon_{\mathrm{rel}}\chi_{\mathbf{ss}}^{\mathbf{K}}\left(E_{\mathbf{K}}\right)$ but larger than $\epsilon_{\mathrm{rel}}\chi_{\mathbf{st}}^{\mathbf{K}}\left(E_{\mathbf{K}}\right)$ in order to lower the computational cost for the same accuracy.  That the eigenvalues of $\chi^{\mathbf{K}}\left(E_{\mathbf{K}}\right)$ are computed to the appropriate relative precision with this increased tolerance for the off-diagonal elements follows from the fact that the maximal difference between eigenvalues of two matrices $A$ and $M$ is bounded by the 2-norm difference of the two matrices,
\begin{align}
\max_j\left|\lambda_j\left[A\right]-\lambda_j\left[M\right]\right|\le \Vert A-M\Vert\, ,
\end{align}
with $\lambda_j\left[A\right]$ the $j^{\mathrm{th}}$ eigenvalue of $A$~\cite{golubGH1996}.

\section{Solution of the nonlinear eigenvalue problem}
\label{sec:NLEEsoln}

In this section we focus on the numerical solution of the nonlinear eigenvalue equation Eq.~\eqref{eq:NLEE}.  For simplicity, we will focus on the limit of an infinitely broad resonance in which $\tilde{g}\to\infty$, $h\nu/E_R\to\infty$, $\tilde{a}_s\to$const such that Eq.~\eqref{eq:NLEE} becomes
\begin{align}
\label{eq:infbroadNLEE}\frac{\pi}{8}\frac{1}{\tilde{a}_s}\Upsilon_{\mathbf{s}}^{\mathbf{K}}&=\sum_{\mathbf{t}}\chi_{\mathbf{st}}^{\mathbf{K}}\left(E_{\mathbf{K}}/E_R\right)\Upsilon_{\mathbf{t}}^{\mathbf{K}}\, .
\end{align}
This is essentially the limit of a zero-range resonance, $r_B/a\to 0$ in which the $s$-wave scattering length is the only relevant parameter.  The method here also applies to finite width resonances~\cite{wall_carr_12}.

The simplest means to solve Eq.~\eqref{eq:infbroadNLEE} is to fix the eigenvalue $E_{\mathbf{K}}$ and then solve Eq.~\eqref{eq:infbroadNLEE} as a linear eigenvalue problem for $\tilde{a}_s$.  We will refer to this method of finding solutions as the linear method.  The computation of Hubbard parameters requires the bound state solution in the entire total quasimomentum BZ at fixed $\tilde{a}_s$ rather than fixed $E_{\mathbf{K}}$, see Eq.~\eqref{eq:effpairing}, for example.

To solve Eq.~\eqref{eq:infbroadNLEE} for fixed $\tilde{a}_s$, we first restate the problem as as
\begin{align}
T_{\mathbf{K};\tilde{a}_s}\left(E_{\alpha}^{\mathbf{K}}\right)\boldsymbol{\Upsilon}^{\mathbf{K}}_{\alpha}=\left[\chi^{\mathbf{K}}\left({E_{\mathbf{K}}}/{E_R}\right)-\frac{\pi}{8}\frac{1}{\tilde{a}_s}\hat{1}\right]\boldsymbol{\Upsilon}^{\mathbf{K}}_{\alpha}&=0
\end{align}
We assume that we have some guess at the eigenenergy $E_{\mathbf{K}}$, which will not necessarily be the solution but will sit on some continuous path traced by an eigenvalue $\lambda_{\mathbf{K}\tilde{a}_s\alpha}\left(E_{\mathbf{K}}\right)$ of $T^{\mathbf{K};\tilde{a}_s}\left(E_{\mathbf{K}}\right)$ provided that the guess does not lie in one of the scattering bands of the open channel~\cite{lancaster_66}.  Hence, we can perform a root-finding method on $\lambda_{\mathbf{K}\tilde{a}_s\alpha}\left(E_{\mathbf{K}}\right)$ to search for exact eigentuples.  The Newton iteration yields the shift in the approximate eigenvalue $\tilde{E}^{\alpha}_{\mathbf{K}}/E_R$ at the $i^{th}$ iteration as
\begin{align}
\label{eq:NRiter}s_{i}&=-\frac{\tilde{\boldsymbol{\Upsilon}}^{\mathbf{K}}_{\alpha}\cdot T_{\mathbf{K};\tilde{a}_s}\left(\tilde{E}^{\alpha}_{\mathbf{K}}\right)\cdot \tilde{\boldsymbol{\Upsilon}}^{\mathbf{K}}_{\alpha}}{\tilde{\boldsymbol{\Upsilon}}^{\mathbf{K}}_{\alpha}\cdot {T'}_{\mathbf{K};\tilde{a}_s}\left(\tilde{E}^{\alpha}_{\mathbf{K}}\right)\cdot \tilde{\boldsymbol{\Upsilon}}^{\mathbf{K}}_{\alpha}}\, ,
\end{align}
where $\tilde{\boldsymbol{\Upsilon}}^{\mathbf{K}}_{\alpha}$ is the eigenvector of $T_{\mathbf{K};\tilde{a}_s}\left(\tilde{E}^{\alpha}_{\mathbf{K}}\right)$ which is smoothly connected to our initial guess.  As $\tilde{a}_s$ is a fixed parameter appearing in $T_{\mathbf{K}\tilde{a}_s}\left(E_{\mathbf{K}}\right)$, the denominator is in fact the quantity $\tilde{\boldsymbol{\Upsilon}}^{\mathbf{K}}_{\alpha}\cdot\chi'\left(E_{\mathbf{K}}^{\alpha}/E_R\right)\cdot\tilde{\boldsymbol{\Upsilon}}^{\mathbf{K}}_{\alpha}$ appearing in the normalization factor of the two-particle solution.  One updates the guess at the eigenenergy using this shift and iteration continues until the shift drops below a given tolerance.  This \emph{Newton-Raphson} method can be proven to be quadratically convergent locally~\cite{lancaster_66}.

The complete procedure for finding the band structure at fixed $\tilde{a}_s$ is first to find the exact solutions $\left(\tilde{a}_s,E_{\mathbf{K}},\boldsymbol{\Upsilon}^{\mathbf{K}}\right)$ at the extremal points of the BZ where all elements of $\mathbf{K}$ are either $0$ or $-\pi/a$.  This is done by choosing a range of $E_{\mathbf{K}}$ and solving the associated linear eigenproblem for $\tilde{a}_s$ by the linear method.  As discussed in Sec.~\ref{sec:Wannier}, these solutions can be classified according to their parity unambiguously.  Now, $\tilde{a}_s$ is fixed and the solutions at the extremal points are used as guesses for solutions at fixed $\tilde{a}_s$ and $\mathbf{K}$ with a desired parity.  A nice feature of the Newton-Raphson iteration Eq.~\eqref{eq:NRiter} is that we can search along directions which are smoothly connected to guesses with definite parity, and so the Kohn construction discussed in Sec.~\ref{sec:Wannier} can be enforced.

{\bibliographystyle{unsrtnat}}
{\bibliography{running}}

\begin{thebibliography}{70}
\expandafter\ifx\csname natexlab\endcsname\relax\def\natexlab#1{#1}\fi
\expandafter\ifx\csname bibnamefont\endcsname\relax
  \def\bibnamefont#1{#1}\fi
\expandafter\ifx\csname bibfnamefont\endcsname\relax
  \def\bibfnamefont#1{#1}\fi
\expandafter\ifx\csname citenamefont\endcsname\relax
  \def\citenamefont#1{#1}\fi
\expandafter\ifx\csname url\endcsname\relax
  \def\url#1{\texttt{#1}}\fi
\expandafter\ifx\csname urlprefix\endcsname\relax\def\urlprefix{URL }\fi
\providecommand{\bibinfo}[2]{#2}
\providecommand{\eprint}[2][]{\url{#2}}

\bibitem[{\citenamefont{Leggett}(2006)}]{Leggett_06}
\bibinfo{author}{\bibfnamefont{A.~J.} \bibnamefont{Leggett}},
  \emph{\bibinfo{title}{Quantum liquids: {B}ose condensation and {C}ooper
  pairing in condensed-matter systems}} (\bibinfo{publisher}{Oxford University
  Press}, \bibinfo{address}{Oxford}, \bibinfo{year}{2006}).

\bibitem[{\citenamefont{Leggett}(1980)}]{Leggett_80}
\bibinfo{author}{\bibfnamefont{A.~J.} \bibnamefont{Leggett}}, in
  \emph{\bibinfo{booktitle}{Modern {T}rends in the {T}heory of {C}ondensed
  {M}atter}}, edited by
  \bibinfo{editor}{\bibfnamefont{A.}~\bibnamefont{Pekalski}} \bibnamefont{and}
  \bibinfo{editor}{\bibfnamefont{R.}~\bibnamefont{Przystawa}}
  (\bibinfo{publisher}{Springer-Verlag}, \bibinfo{address}{Berlin},
  \bibinfo{year}{1980}).

\bibitem[{\citenamefont{Eagles}(1969)}]{Eagles_69}
\bibinfo{author}{\bibfnamefont{D.~M.} \bibnamefont{Eagles}},
  \bibinfo{journal}{Phys. Rev.} \textbf{\bibinfo{volume}{186}},
  \bibinfo{pages}{456} (\bibinfo{year}{1969}).

\bibitem[{\citenamefont{Nozi{\' e}res and
  Schmitt-Rink}(1985)}]{Nozieres_SchmittRink_85}
\bibinfo{author}{\bibfnamefont{P.}~\bibnamefont{Nozi{\' e}res}}
  \bibnamefont{and}
  \bibinfo{author}{\bibfnamefont{S.}~\bibnamefont{Schmitt-Rink}},
  \bibinfo{journal}{J.~Low Temp.~Phys.} \textbf{\bibinfo{volume}{59}},
  \bibinfo{pages}{195} (\bibinfo{year}{1985}), ISSN \bibinfo{issn}{0022-2291},
  \bibinfo{note}{10.1007/BF00683774},
  \urlprefix\url{http://dx.doi.org/10.1007/BF00683774}.

\bibitem[{\citenamefont{Chen et~al.}(2005)\citenamefont{Chen, Stajic, Tan, and
  Levin}}]{Chen_Stajic_05}
\bibinfo{author}{\bibfnamefont{Q.}~\bibnamefont{Chen}},
  \bibinfo{author}{\bibfnamefont{J.}~\bibnamefont{Stajic}},
  \bibinfo{author}{\bibfnamefont{S.}~\bibnamefont{Tan}}, \bibnamefont{and}
  \bibinfo{author}{\bibfnamefont{K.}~\bibnamefont{Levin}},
  \bibinfo{journal}{Physics Reports} \textbf{\bibinfo{volume}{412}},
  \bibinfo{pages}{1 } (\bibinfo{year}{2005}), ISSN \bibinfo{issn}{0370-1573},
  \urlprefix\url{http://www.sciencedirect.com/science/article/pii/S0370157305001067}.

\bibitem[{\citenamefont{Regal et~al.}(2004{\natexlab{a}})\citenamefont{Regal,
  Greiner, and Jin}}]{Regal_Greiner_04b}
\bibinfo{author}{\bibfnamefont{C.~A.} \bibnamefont{Regal}},
  \bibinfo{author}{\bibfnamefont{M.}~\bibnamefont{Greiner}}, \bibnamefont{and}
  \bibinfo{author}{\bibfnamefont{D.~S.} \bibnamefont{Jin}},
  \bibinfo{journal}{Phys. Rev. Lett.} \textbf{\bibinfo{volume}{92}},
  \bibinfo{pages}{040403} (\bibinfo{year}{2004}{\natexlab{a}}),
  \urlprefix\url{http://link.aps.org/doi/10.1103/PhysRevLett.92.040403}.

\bibitem[{\citenamefont{Zwierlein et~al.}(2004)\citenamefont{Zwierlein, Stan,
  Schunck, Raupach, Kerman, and Ketterle}}]{Zwierlein_Stan_04}
\bibinfo{author}{\bibfnamefont{M.~W.} \bibnamefont{Zwierlein}},
  \bibinfo{author}{\bibfnamefont{C.~A.} \bibnamefont{Stan}},
  \bibinfo{author}{\bibfnamefont{C.~H.} \bibnamefont{Schunck}},
  \bibinfo{author}{\bibfnamefont{S.~M.~F.} \bibnamefont{Raupach}},
  \bibinfo{author}{\bibfnamefont{A.~J.} \bibnamefont{Kerman}},
  \bibnamefont{and} \bibinfo{author}{\bibfnamefont{W.}~\bibnamefont{Ketterle}},
  \bibinfo{journal}{Phys. Rev. Lett.} \textbf{\bibinfo{volume}{92}},
  \bibinfo{pages}{120403} (\bibinfo{year}{2004}).

\bibitem[{\citenamefont{Bartenstein et~al.}(2004)\citenamefont{Bartenstein,
  Altmeyer, Riedl, Jochim, Chin, Denschlag, and
  Grimm}}]{Barteinstein_Altmeyer_04}
\bibinfo{author}{\bibfnamefont{M.}~\bibnamefont{Bartenstein}},
  \bibinfo{author}{\bibfnamefont{A.}~\bibnamefont{Altmeyer}},
  \bibinfo{author}{\bibfnamefont{S.}~\bibnamefont{Riedl}},
  \bibinfo{author}{\bibfnamefont{S.}~\bibnamefont{Jochim}},
  \bibinfo{author}{\bibfnamefont{C.}~\bibnamefont{Chin}},
  \bibinfo{author}{\bibfnamefont{J.~H.} \bibnamefont{Denschlag}},
  \bibnamefont{and} \bibinfo{author}{\bibfnamefont{R.}~\bibnamefont{Grimm}},
  \bibinfo{journal}{Phys. Rev. Lett.} \textbf{\bibinfo{volume}{92}},
  \bibinfo{pages}{120401} (\bibinfo{year}{2004}).

\bibitem[{\citenamefont{Kinast et~al.}(2004)\citenamefont{Kinast, Hemmer, Gehm,
  Turlapov, and Thomas}}]{Kinast_Hemmer_04}
\bibinfo{author}{\bibfnamefont{J.}~\bibnamefont{Kinast}},
  \bibinfo{author}{\bibfnamefont{S.~L.} \bibnamefont{Hemmer}},
  \bibinfo{author}{\bibfnamefont{M.~E.} \bibnamefont{Gehm}},
  \bibinfo{author}{\bibfnamefont{A.}~\bibnamefont{Turlapov}}, \bibnamefont{and}
  \bibinfo{author}{\bibfnamefont{J.~E.} \bibnamefont{Thomas}},
  \bibinfo{journal}{Phys. Rev. Lett.} \textbf{\bibinfo{volume}{92}},
  \bibinfo{pages}{150402} (\bibinfo{year}{2004}).

\bibitem[{\citenamefont{Bourdel et~al.}(2004)\citenamefont{Bourdel, Khaykovich,
  Cubizolles, Zhang, Chevy, Teichmann, Tarruell, Kokkelmans, and
  Salomon}}]{Bourdel_Khaykovich_04}
\bibinfo{author}{\bibfnamefont{T.}~\bibnamefont{Bourdel}},
  \bibinfo{author}{\bibfnamefont{L.}~\bibnamefont{Khaykovich}},
  \bibinfo{author}{\bibfnamefont{J.}~\bibnamefont{Cubizolles}},
  \bibinfo{author}{\bibfnamefont{J.}~\bibnamefont{Zhang}},
  \bibinfo{author}{\bibfnamefont{F.}~\bibnamefont{Chevy}},
  \bibinfo{author}{\bibfnamefont{M.}~\bibnamefont{Teichmann}},
  \bibinfo{author}{\bibfnamefont{L.}~\bibnamefont{Tarruell}},
  \bibinfo{author}{\bibfnamefont{S.~J. J. M.~F.} \bibnamefont{Kokkelmans}},
  \bibnamefont{and} \bibinfo{author}{\bibfnamefont{C.}~\bibnamefont{Salomon}},
  \bibinfo{journal}{Phys. Rev. Lett.} \textbf{\bibinfo{volume}{93}},
  \bibinfo{pages}{050401} (\bibinfo{year}{2004}).

\bibitem[{\citenamefont{Radzihovsky et~al.}(2004)\citenamefont{Radzihovsky,
  Park, and Weichman}}]{Radzihovsky_Park_04}
\bibinfo{author}{\bibfnamefont{L.}~\bibnamefont{Radzihovsky}},
  \bibinfo{author}{\bibfnamefont{J.}~\bibnamefont{Park}}, \bibnamefont{and}
  \bibinfo{author}{\bibfnamefont{P.~B.} \bibnamefont{Weichman}},
  \bibinfo{journal}{Phys. Rev. Lett.} \textbf{\bibinfo{volume}{92}},
  \bibinfo{pages}{160402} (\bibinfo{year}{2004}),
  \urlprefix\url{http://link.aps.org/doi/10.1103/PhysRevLett.92.160402}.

\bibitem[{\citenamefont{Radzihovsky et~al.}(2008)\citenamefont{Radzihovsky,
  Weichman, and Park}}]{Radzihovsky_Weichman_08}
\bibinfo{author}{\bibfnamefont{L.}~\bibnamefont{Radzihovsky}},
  \bibinfo{author}{\bibfnamefont{P.~B.} \bibnamefont{Weichman}},
  \bibnamefont{and} \bibinfo{author}{\bibfnamefont{J.~I.} \bibnamefont{Park}},
  \bibinfo{journal}{Annals of Physics} \textbf{\bibinfo{volume}{323}},
  \bibinfo{pages}{2376 } (\bibinfo{year}{2008}), ISSN
  \bibinfo{issn}{0003-4916},
  \urlprefix\url{http://www.sciencedirect.com/science/article/pii/S0003491608000742}.

\bibitem[{\citenamefont{Ejima et~al.}(2011)\citenamefont{Ejima, Bhaseen,
  Hohenadler, Essler, Fehske, and Simons}}]{Ejima_Bhaseen_11}
\bibinfo{author}{\bibfnamefont{S.}~\bibnamefont{Ejima}},
  \bibinfo{author}{\bibfnamefont{M.~J.} \bibnamefont{Bhaseen}},
  \bibinfo{author}{\bibfnamefont{M.}~\bibnamefont{Hohenadler}},
  \bibinfo{author}{\bibfnamefont{F.~H.~L.} \bibnamefont{Essler}},
  \bibinfo{author}{\bibfnamefont{H.}~\bibnamefont{Fehske}}, \bibnamefont{and}
  \bibinfo{author}{\bibfnamefont{B.~D.} \bibnamefont{Simons}},
  \bibinfo{journal}{Phys. Rev. Lett.} \textbf{\bibinfo{volume}{106}},
  \bibinfo{pages}{015303} (\bibinfo{year}{2011}).

\bibitem[{\citenamefont{Shankar}(1994)}]{Shankar_94}
\bibinfo{author}{\bibfnamefont{R.}~\bibnamefont{Shankar}},
  \bibinfo{journal}{Rev. Mod. Phys.} \textbf{\bibinfo{volume}{66}},
  \bibinfo{pages}{129} (\bibinfo{year}{1994}),
  \urlprefix\url{http://link.aps.org/doi/10.1103/RevModPhys.66.129}.

\bibitem[{\citenamefont{Wall and Carr}(2012)}]{wall_carr_12}
\bibinfo{author}{\bibfnamefont{M.~L.} \bibnamefont{Wall}} \bibnamefont{and}
  \bibinfo{author}{\bibfnamefont{L.~D.} \bibnamefont{Carr}},
  \bibinfo{journal}{Phys. Rev. Lett.} \textbf{\bibinfo{volume}{109}},
  \bibinfo{pages}{055302} (\bibinfo{year}{2012}),
  \urlprefix\url{http://link.aps.org/doi/10.1103/PhysRevLett.109.055302}.

\bibitem[{\citenamefont{Andreev et~al.}(2004)\citenamefont{Andreev, Gurarie,
  and Radzihovsky}}]{Andreev_Gurarie_04}
\bibinfo{author}{\bibfnamefont{A.~V.} \bibnamefont{Andreev}},
  \bibinfo{author}{\bibfnamefont{V.}~\bibnamefont{Gurarie}}, \bibnamefont{and}
  \bibinfo{author}{\bibfnamefont{L.}~\bibnamefont{Radzihovsky}},
  \bibinfo{journal}{Phys. Rev. Lett.} \textbf{\bibinfo{volume}{93}},
  \bibinfo{pages}{130402} (\bibinfo{year}{2004}),
  \urlprefix\url{http://link.aps.org/doi/10.1103/PhysRevLett.93.130402}.

\bibitem[{\citenamefont{Gurarie and
  Radzihovsky}(2007)}]{Gurarie_Radzihovsky_07}
\bibinfo{author}{\bibfnamefont{V.}~\bibnamefont{Gurarie}} \bibnamefont{and}
  \bibinfo{author}{\bibfnamefont{L.}~\bibnamefont{Radzihovsky}},
  \bibinfo{journal}{Ann.~Phys.} \textbf{\bibinfo{volume}{322}},
  \bibinfo{pages}{2 } (\bibinfo{year}{2007}), ISSN \bibinfo{issn}{0003-4916},
  \bibinfo{note}{january Special Issue 2007},
  \urlprefix\url{http://www.sciencedirect.com/science/article/pii/S0003491606002399}.

\bibitem[{\citenamefont{Timmermans et~al.}(2001)\citenamefont{Timmermans,
  Furuya, Milonni, and Kerman}}]{Timmermans_Furuya_01}
\bibinfo{author}{\bibfnamefont{E.}~\bibnamefont{Timmermans}},
  \bibinfo{author}{\bibfnamefont{K.}~\bibnamefont{Furuya}},
  \bibinfo{author}{\bibfnamefont{P.~W.} \bibnamefont{Milonni}},
  \bibnamefont{and} \bibinfo{author}{\bibfnamefont{A.~K.}
  \bibnamefont{Kerman}}, \bibinfo{journal}{Phys. Lett. A}
  \textbf{\bibinfo{volume}{285}}, \bibinfo{pages}{228} (\bibinfo{year}{2001}).

\bibitem[{\citenamefont{Holland et~al.}(2001)\citenamefont{Holland, Kokkelmans,
  Chiofalo, and Walser}}]{Holland_Kokkelmans_01}
\bibinfo{author}{\bibfnamefont{M.}~\bibnamefont{Holland}},
  \bibinfo{author}{\bibfnamefont{S.~J. J. M.~F.} \bibnamefont{Kokkelmans}},
  \bibinfo{author}{\bibfnamefont{M.~L.} \bibnamefont{Chiofalo}},
  \bibnamefont{and} \bibinfo{author}{\bibfnamefont{R.}~\bibnamefont{Walser}},
  \bibinfo{journal}{Phys. Rev. Lett.} \textbf{\bibinfo{volume}{87}},
  \bibinfo{pages}{120406} (\bibinfo{year}{2001}).

\bibitem[{\citenamefont{Kokkelmans et~al.}(2002)\citenamefont{Kokkelmans,
  Milstein, Chiofalo, Walser, and Holland}}]{Kokkelmans_Milstein_02}
\bibinfo{author}{\bibfnamefont{S.~J. J. M.~F.} \bibnamefont{Kokkelmans}},
  \bibinfo{author}{\bibfnamefont{J.~N.} \bibnamefont{Milstein}},
  \bibinfo{author}{\bibfnamefont{M.~L.} \bibnamefont{Chiofalo}},
  \bibinfo{author}{\bibfnamefont{R.}~\bibnamefont{Walser}}, \bibnamefont{and}
  \bibinfo{author}{\bibfnamefont{M.~J.} \bibnamefont{Holland}},
  \bibinfo{journal}{Phys. Rev. A} \textbf{\bibinfo{volume}{65}},
  \bibinfo{pages}{053617} (\bibinfo{year}{2002}).

\bibitem[{\citenamefont{Milstein et~al.}(2002)\citenamefont{Milstein,
  Kokkelmans, and Holland}}]{Milstein_Kokkelmans_02}
\bibinfo{author}{\bibfnamefont{J.~N.} \bibnamefont{Milstein}},
  \bibinfo{author}{\bibfnamefont{S.~J. J. M.~F.} \bibnamefont{Kokkelmans}},
  \bibnamefont{and} \bibinfo{author}{\bibfnamefont{M.~J.}
  \bibnamefont{Holland}}, \bibinfo{journal}{Phys. Rev. A}
  \textbf{\bibinfo{volume}{66}}, \bibinfo{pages}{043604}
  (\bibinfo{year}{2002}).

\bibitem[{\citenamefont{Ohashi and Griffin}(2002)}]{Ohashi_Griffin_02}
\bibinfo{author}{\bibfnamefont{Y.}~\bibnamefont{Ohashi}} \bibnamefont{and}
  \bibinfo{author}{\bibfnamefont{A.}~\bibnamefont{Griffin}},
  \bibinfo{journal}{Phys. Rev. Lett.} \textbf{\bibinfo{volume}{89}},
  \bibinfo{pages}{130402} (\bibinfo{year}{2002}).

\bibitem[{\citenamefont{Stajic et~al.}(2004)\citenamefont{Stajic, Milstein,
  Chen, Chiofalo, Holland, and Levin}}]{Stajic_Milstein_04}
\bibinfo{author}{\bibfnamefont{J.}~\bibnamefont{Stajic}},
  \bibinfo{author}{\bibfnamefont{J.~N.} \bibnamefont{Milstein}},
  \bibinfo{author}{\bibfnamefont{Q.}~\bibnamefont{Chen}},
  \bibinfo{author}{\bibfnamefont{M.~L.} \bibnamefont{Chiofalo}},
  \bibinfo{author}{\bibfnamefont{M.~J.} \bibnamefont{Holland}},
  \bibnamefont{and} \bibinfo{author}{\bibfnamefont{K.}~\bibnamefont{Levin}},
  \bibinfo{journal}{Phys. Rev. A} \textbf{\bibinfo{volume}{69}},
  \bibinfo{pages}{063610} (\bibinfo{year}{2004}).

\bibitem[{\citenamefont{Timmermans et~al.}(1999)\citenamefont{Timmermans,
  Tommasini, Hussein, and Kerman}}]{Timmermans_Tommasini_99}
\bibinfo{author}{\bibfnamefont{E.}~\bibnamefont{Timmermans}},
  \bibinfo{author}{\bibfnamefont{P.}~\bibnamefont{Tommasini}},
  \bibinfo{author}{\bibfnamefont{M.}~\bibnamefont{Hussein}}, \bibnamefont{and}
  \bibinfo{author}{\bibfnamefont{A.}~\bibnamefont{Kerman}},
  \bibinfo{journal}{Phys. Rep.} \textbf{\bibinfo{volume}{315}},
  \bibinfo{pages}{199 } (\bibinfo{year}{1999}), ISSN \bibinfo{issn}{0370-1573},
  \urlprefix\url{http://www.sciencedirect.com/science/article/pii/S0370157399000253}.

\bibitem[{\citenamefont{Ranninger and
  Robaszkiewicz}(1985)}]{Ranninger_Robaszkiewicz_85}
\bibinfo{author}{\bibfnamefont{J.}~\bibnamefont{Ranninger}} \bibnamefont{and}
  \bibinfo{author}{\bibfnamefont{S.}~\bibnamefont{Robaszkiewicz}},
  \bibinfo{journal}{Physica B} \textbf{\bibinfo{volume}{135}},
  \bibinfo{pages}{468} (\bibinfo{year}{1985}).

\bibitem[{\citenamefont{Friedberg and Lee}(1989)}]{Friedberg_Lee_89}
\bibinfo{author}{\bibfnamefont{R.}~\bibnamefont{Friedberg}} \bibnamefont{and}
  \bibinfo{author}{\bibfnamefont{T.~D.} \bibnamefont{Lee}},
  \bibinfo{journal}{Phys. Rev. B} \textbf{\bibinfo{volume}{40}},
  \bibinfo{pages}{6745} (\bibinfo{year}{1989}).

\bibitem[{\citenamefont{Ranninger and Robin}(1995)}]{Ranninger_Robin_95}
\bibinfo{author}{\bibfnamefont{J.}~\bibnamefont{Ranninger}} \bibnamefont{and}
  \bibinfo{author}{\bibfnamefont{J.~M.} \bibnamefont{Robin}},
  \bibinfo{journal}{Physica C} \textbf{\bibinfo{volume}{253}},
  \bibinfo{pages}{279} (\bibinfo{year}{1995}).

\bibitem[{\citenamefont{Micnas}(2007)}]{Micnas_07}
\bibinfo{author}{\bibfnamefont{R.}~\bibnamefont{Micnas}},
  \bibinfo{journal}{Phys. Rev. B} \textbf{\bibinfo{volume}{76}},
  \bibinfo{pages}{184507} (\bibinfo{year}{2007}),
  \urlprefix\url{http://link.aps.org/doi/10.1103/PhysRevB.76.184507}.

\bibitem[{\citenamefont{Yang et~al.}(2011)\citenamefont{Yang, Kozik, Wang, and
  Troyer}}]{Yang_Kozik_11}
\bibinfo{author}{\bibfnamefont{K.-Y.} \bibnamefont{Yang}},
  \bibinfo{author}{\bibfnamefont{E.}~\bibnamefont{Kozik}},
  \bibinfo{author}{\bibfnamefont{X.}~\bibnamefont{Wang}}, \bibnamefont{and}
  \bibinfo{author}{\bibfnamefont{M.}~\bibnamefont{Troyer}},
  \bibinfo{journal}{Phys. Rev. B} \textbf{\bibinfo{volume}{83}},
  \bibinfo{pages}{214516} (\bibinfo{year}{2011}),
  \urlprefix\url{http://link.aps.org/doi/10.1103/PhysRevB.83.214516}.

\bibitem[{\citenamefont{Holland et~al.}(2005)\citenamefont{Holland, Menotti,
  and Viverit}}]{HollandProceedings}
\bibinfo{author}{\bibfnamefont{M.}~\bibnamefont{Holland}},
  \bibinfo{author}{\bibfnamefont{C.}~\bibnamefont{Menotti}}, \bibnamefont{and}
  \bibinfo{author}{\bibfnamefont{L.}~\bibnamefont{Viverit}},
  \bibinfo{journal}{AIP Conf. Proc.} \textbf{\bibinfo{volume}{770}},
  \bibinfo{pages}{238} (\bibinfo{year}{2005}).

\bibitem[{\citenamefont{Holland et~al.}(2004)\citenamefont{Holland, Menotti,
  and Viverit}}]{HMV}
\bibinfo{author}{\bibfnamefont{M.~J.} \bibnamefont{Holland}},
  \bibinfo{author}{\bibfnamefont{C.}~\bibnamefont{Menotti}}, \bibnamefont{and}
  \bibinfo{author}{\bibfnamefont{L.}~\bibnamefont{Viverit}},
  \bibinfo{journal}{http://arxiv.org/abs/cond-mat/0404234v2}
  (\bibinfo{year}{2004}).

\bibitem[{\citenamefont{Esslinger}(2010)}]{Esslinger_10}
\bibinfo{author}{\bibfnamefont{T.}~\bibnamefont{Esslinger}},
  \bibinfo{journal}{Annu.~Rev.~Condens.~Matter Phys.}
  \textbf{\bibinfo{volume}{1}}, \bibinfo{pages}{129} (\bibinfo{year}{2010}),
  \eprint{http://www.annualreviews.org/doi/pdf/10.1146/annurev-conmatphys-070909-104059},
  \urlprefix\url{http://www.annualreviews.org/doi/abs/10.1146/annurev-conmatphys-070909-104059}.

\bibitem[{\citenamefont{Carr and Holland}(2005)}]{Carr_Holland_05}
\bibinfo{author}{\bibfnamefont{L.~D.} \bibnamefont{Carr}} \bibnamefont{and}
  \bibinfo{author}{\bibfnamefont{M.~J.} \bibnamefont{Holland}},
  \bibinfo{journal}{Phys. Rev. A} \textbf{\bibinfo{volume}{72}},
  \bibinfo{pages}{033622(R)} (\bibinfo{year}{2005}).

\bibitem[{\citenamefont{Zhou}(2005)}]{Zhou_05}
\bibinfo{author}{\bibfnamefont{F.}~\bibnamefont{Zhou}}, \bibinfo{journal}{Phys.
  Rev. B} \textbf{\bibinfo{volume}{72}}, \bibinfo{pages}{220501}
  (\bibinfo{year}{2005}),
  \urlprefix\url{http://link.aps.org/doi/10.1103/PhysRevB.72.220501}.

\bibitem[{\citenamefont{Dickerscheid et~al.}(2005)\citenamefont{Dickerscheid,
  Khawaja, {van Oosten}, and Stoof}}]{Dickerscheid_Khawaja_05}
\bibinfo{author}{\bibfnamefont{D.~B.~M.} \bibnamefont{Dickerscheid}},
  \bibinfo{author}{\bibfnamefont{U.~A.} \bibnamefont{Khawaja}},
  \bibinfo{author}{\bibfnamefont{D.}~\bibnamefont{{van Oosten}}},
  \bibnamefont{and} \bibinfo{author}{\bibfnamefont{H.~T.~C.}
  \bibnamefont{Stoof}}, \bibinfo{journal}{Phys. Rev. A}
  \textbf{\bibinfo{volume}{71}}, \bibinfo{pages}{043604}
  (\bibinfo{year}{2005}).

\bibitem[{\citenamefont{Diener and Ho}(2006)}]{Diener_Ho_06}
\bibinfo{author}{\bibfnamefont{R.~B.} \bibnamefont{Diener}} \bibnamefont{and}
  \bibinfo{author}{\bibfnamefont{T.-L.} \bibnamefont{Ho}},
  \bibinfo{journal}{Phys. Rev. Lett.} \textbf{\bibinfo{volume}{96}},
  \bibinfo{pages}{010402} (\bibinfo{year}{2006}).

\bibitem[{\citenamefont{Duan}(2005)}]{Duan_05}
\bibinfo{author}{\bibfnamefont{L.-M.} \bibnamefont{Duan}},
  \bibinfo{journal}{Phys. Rev. Lett.} \textbf{\bibinfo{volume}{95}},
  \bibinfo{eid}{243202} (pages~\bibinfo{numpages}{4}) (\bibinfo{year}{2005}).

\bibitem[{\citenamefont{Duan}(2008)}]{Duan_08}
\bibinfo{author}{\bibfnamefont{L.-M.} \bibnamefont{Duan}},
  \bibinfo{journal}{Europhys.~Lett.} \textbf{\bibinfo{volume}{81}},
  \bibinfo{pages}{20001} (\bibinfo{year}{2008}),
  \urlprefix\url{http://stacks.iop.org/0295-5075/81/i=2/a=20001}.

\bibitem[{\citenamefont{Kestner and Duan}(2010)}]{Kestner_Duan_10}
\bibinfo{author}{\bibfnamefont{J.~P.} \bibnamefont{Kestner}} \bibnamefont{and}
  \bibinfo{author}{\bibfnamefont{L.-M.} \bibnamefont{Duan}},
  \bibinfo{journal}{Phys. Rev. A} \textbf{\bibinfo{volume}{81}},
  \bibinfo{pages}{043618} (\bibinfo{year}{2010}).

\bibitem[{\citenamefont{Grishkevich et~al.}(2011)\citenamefont{Grishkevich,
  Sala, and Saenz}}]{Grishkevich_Sala_11}
\bibinfo{author}{\bibfnamefont{S.}~\bibnamefont{Grishkevich}},
  \bibinfo{author}{\bibfnamefont{S.}~\bibnamefont{Sala}}, \bibnamefont{and}
  \bibinfo{author}{\bibfnamefont{A.}~\bibnamefont{Saenz}},
  \bibinfo{journal}{Phys. Rev. A} \textbf{\bibinfo{volume}{84}},
  \bibinfo{pages}{062710} (\bibinfo{year}{2011}),
  \urlprefix\url{http://link.aps.org/doi/10.1103/PhysRevA.84.062710}.

\bibitem[{\citenamefont{Mentink and Kokkelmans}(2009)}]{Mentink_Kokkelmans_09}
\bibinfo{author}{\bibfnamefont{J.}~\bibnamefont{Mentink}} \bibnamefont{and}
  \bibinfo{author}{\bibfnamefont{S.}~\bibnamefont{Kokkelmans}},
  \bibinfo{journal}{Phys. Rev. A} \textbf{\bibinfo{volume}{79}},
  \bibinfo{pages}{032709} (\bibinfo{year}{2009}),
  \urlprefix\url{http://link.aps.org/doi/10.1103/PhysRevA.79.032709}.

\bibitem[{\citenamefont{B\"uchler}(2010)}]{Buechler_10}
\bibinfo{author}{\bibfnamefont{H.~P.} \bibnamefont{B\"uchler}},
  \bibinfo{journal}{Phys. Rev. Lett.} \textbf{\bibinfo{volume}{104}},
  \bibinfo{pages}{090402} (\bibinfo{year}{2010}).

\bibitem[{\citenamefont{B\"uchler}(2012)}]{Buechler_12}
\bibinfo{author}{\bibfnamefont{H.~P.} \bibnamefont{B\"uchler}},
  \bibinfo{journal}{Phys. Rev. Lett.} \textbf{\bibinfo{volume}{108}},
  \bibinfo{pages}{069903} (\bibinfo{year}{2012}),
  \urlprefix\url{http://link.aps.org/doi/10.1103/PhysRevLett.108.069903}.

\bibitem[{\citenamefont{von Stecher et~al.}(2011)\citenamefont{von Stecher,
  Gurarie, Radzihovsky, and Rey}}]{von_Stecher_Gurarie_11}
\bibinfo{author}{\bibfnamefont{J.}~\bibnamefont{von Stecher}},
  \bibinfo{author}{\bibfnamefont{V.}~\bibnamefont{Gurarie}},
  \bibinfo{author}{\bibfnamefont{L.}~\bibnamefont{Radzihovsky}},
  \bibnamefont{and} \bibinfo{author}{\bibfnamefont{A.~M.} \bibnamefont{Rey}},
  \bibinfo{journal}{Phys. Rev. Lett.} \textbf{\bibinfo{volume}{106}},
  \bibinfo{pages}{235301} (\bibinfo{year}{2011}).

\bibitem[{\citenamefont{Titvinidze et~al.}(2010)\citenamefont{Titvinidze,
  Snoek, and Hofstetter}}]{Titvinidze_Snoek_11}
\bibinfo{author}{\bibfnamefont{I.}~\bibnamefont{Titvinidze}},
  \bibinfo{author}{\bibfnamefont{M.}~\bibnamefont{Snoek}}, \bibnamefont{and}
  \bibinfo{author}{\bibfnamefont{W.}~\bibnamefont{Hofstetter}},
  \bibinfo{journal}{New J.~Phys.} \textbf{\bibinfo{volume}{12}},
  \bibinfo{pages}{065030} (\bibinfo{year}{2010}),
  \urlprefix\url{http://stacks.iop.org/1367-2630/12/i=6/a=065030}.

\bibitem[{\citenamefont{Greiner et~al.}(2003)\citenamefont{Greiner, Regal, and
  Jin}}]{greiner2003}
\bibinfo{author}{\bibfnamefont{M.}~\bibnamefont{Greiner}},
  \bibinfo{author}{\bibfnamefont{C.~A.} \bibnamefont{Regal}}, \bibnamefont{and}
  \bibinfo{author}{\bibfnamefont{D.~S.} \bibnamefont{Jin}},
  \bibinfo{journal}{Nature} \textbf{\bibinfo{volume}{426}},
  \bibinfo{pages}{437} (\bibinfo{year}{2003}).

\bibitem[{\citenamefont{Regal et~al.}(2004{\natexlab{b}})\citenamefont{Regal,
  Greiner, and Jin}}]{Regal_Greiner_04}
\bibinfo{author}{\bibfnamefont{C.~A.} \bibnamefont{Regal}},
  \bibinfo{author}{\bibfnamefont{M.}~\bibnamefont{Greiner}}, \bibnamefont{and}
  \bibinfo{author}{\bibfnamefont{D.~S.} \bibnamefont{Jin}},
  \bibinfo{journal}{Phys. Rev. Lett.} \textbf{\bibinfo{volume}{92}},
  \bibinfo{pages}{083201} (\bibinfo{year}{2004}{\natexlab{b}}),
  \urlprefix\url{http://link.aps.org/doi/10.1103/PhysRevLett.92.083201}.

\bibitem[{\citenamefont{Stoferle et~al.}(2006)\citenamefont{Stoferle, Moritz,
  Gunter, Kohl, and Esslinger}}]{stoferle2006}
\bibinfo{author}{\bibfnamefont{T.}~\bibnamefont{Stoferle}},
  \bibinfo{author}{\bibfnamefont{H.}~\bibnamefont{Moritz}},
  \bibinfo{author}{\bibfnamefont{K.}~\bibnamefont{Gunter}},
  \bibinfo{author}{\bibfnamefont{M.}~\bibnamefont{Kohl}}, \bibnamefont{and}
  \bibinfo{author}{\bibfnamefont{T.}~\bibnamefont{Esslinger}},
  \bibinfo{journal}{Phys. Rev. Lett.} \textbf{\bibinfo{volume}{96}},
  \bibinfo{pages}{030401} (\bibinfo{year}{2006}).

\bibitem[{\citenamefont{Strecker et~al.}(2003)\citenamefont{Strecker,
  Partridge, and Hulet}}]{strecker2003}
\bibinfo{author}{\bibfnamefont{K.~E.} \bibnamefont{Strecker}},
  \bibinfo{author}{\bibfnamefont{G.~B.} \bibnamefont{Partridge}},
  \bibnamefont{and} \bibinfo{author}{\bibfnamefont{R.~G.} \bibnamefont{Hulet}},
  \bibinfo{journal}{Phys. Rev. Lett.} \textbf{\bibinfo{volume}{91}},
  \bibinfo{pages}{080406} (\bibinfo{year}{2003}).

\bibitem[{\citenamefont{Cubizolles et~al.}(2003)\citenamefont{Cubizolles,
  Bourdel, Kokkelmans, Shlyapnikov, and Salomon}}]{Cubizolles_Bourdel_03}
\bibinfo{author}{\bibfnamefont{J.}~\bibnamefont{Cubizolles}},
  \bibinfo{author}{\bibfnamefont{T.}~\bibnamefont{Bourdel}},
  \bibinfo{author}{\bibfnamefont{S.~J. J. M.~F.} \bibnamefont{Kokkelmans}},
  \bibinfo{author}{\bibfnamefont{G.~V.} \bibnamefont{Shlyapnikov}},
  \bibnamefont{and} \bibinfo{author}{\bibfnamefont{C.}~\bibnamefont{Salomon}},
  \bibinfo{journal}{Phys. Rev. Lett.} \textbf{\bibinfo{volume}{91}},
  \bibinfo{pages}{240401} (\bibinfo{year}{2003}),
  \urlprefix\url{http://link.aps.org/doi/10.1103/PhysRevLett.91.240401}.

\bibitem[{\citenamefont{Jochim et~al.}(2003{\natexlab{a}})\citenamefont{Jochim,
  Bartenstein, Altmeyer, Hendl, Chin, Denschlag, and
  Grimm}}]{Jochim_Bartenstein_03}
\bibinfo{author}{\bibfnamefont{S.}~\bibnamefont{Jochim}},
  \bibinfo{author}{\bibfnamefont{M.}~\bibnamefont{Bartenstein}},
  \bibinfo{author}{\bibfnamefont{A.}~\bibnamefont{Altmeyer}},
  \bibinfo{author}{\bibfnamefont{G.}~\bibnamefont{Hendl}},
  \bibinfo{author}{\bibfnamefont{C.}~\bibnamefont{Chin}},
  \bibinfo{author}{\bibfnamefont{J.~H.} \bibnamefont{Denschlag}},
  \bibnamefont{and} \bibinfo{author}{\bibfnamefont{R.}~\bibnamefont{Grimm}},
  \bibinfo{journal}{Phys. Rev. Lett.} \textbf{\bibinfo{volume}{91}},
  \bibinfo{pages}{240402} (\bibinfo{year}{2003}{\natexlab{a}}),
  \urlprefix\url{http://link.aps.org/doi/10.1103/PhysRevLett.91.240402}.

\bibitem[{\citenamefont{Jochim et~al.}(2003{\natexlab{b}})\citenamefont{Jochim,
  Bartenstein, Altmeyer, Riedl, Chin, Denschlag, and Grimm}}]{jochim2003b}
\bibinfo{author}{\bibfnamefont{S.}~\bibnamefont{Jochim}},
  \bibinfo{author}{\bibfnamefont{M.}~\bibnamefont{Bartenstein}},
  \bibinfo{author}{\bibfnamefont{A.}~\bibnamefont{Altmeyer}},
  \bibinfo{author}{\bibfnamefont{S.}~\bibnamefont{Riedl}},
  \bibinfo{author}{\bibfnamefont{C.}~\bibnamefont{Chin}},
  \bibinfo{author}{\bibfnamefont{J.~H.} \bibnamefont{Denschlag}},
  \bibnamefont{and} \bibinfo{author}{\bibfnamefont{R.}~\bibnamefont{Grimm}},
  \bibinfo{journal}{Science} \textbf{\bibinfo{volume}{302}},
  \bibinfo{pages}{2102} (\bibinfo{year}{2003}{\natexlab{b}}).

\bibitem[{\citenamefont{Zwierlein et~al.}(2003)\citenamefont{Zwierlein, Stan,
  Schunck, Raupach, Gupta, Hadzibabic, and Ketterle}}]{zwierlein2003}
\bibinfo{author}{\bibfnamefont{M.~W.} \bibnamefont{Zwierlein}},
  \bibinfo{author}{\bibfnamefont{C.~A.} \bibnamefont{Stan}},
  \bibinfo{author}{\bibfnamefont{C.~H.} \bibnamefont{Schunck}},
  \bibinfo{author}{\bibfnamefont{S.~M.~F.} \bibnamefont{Raupach}},
  \bibinfo{author}{\bibfnamefont{S.}~\bibnamefont{Gupta}},
  \bibinfo{author}{\bibfnamefont{Z.}~\bibnamefont{Hadzibabic}},
  \bibnamefont{and} \bibinfo{author}{\bibfnamefont{W.}~\bibnamefont{Ketterle}},
  \bibinfo{journal}{Phys. Rev. Lett.} \textbf{\bibinfo{volume}{91}},
  \bibinfo{pages}{250401} (\bibinfo{year}{2003}).

\bibitem[{\citenamefont{Chin et~al.}(2010)\citenamefont{Chin, Grimm, Julienne,
  and Tiesinga}}]{Chin_Grimm_10}
\bibinfo{author}{\bibfnamefont{C.}~\bibnamefont{Chin}},
  \bibinfo{author}{\bibfnamefont{R.}~\bibnamefont{Grimm}},
  \bibinfo{author}{\bibfnamefont{P.}~\bibnamefont{Julienne}}, \bibnamefont{and}
  \bibinfo{author}{\bibfnamefont{E.}~\bibnamefont{Tiesinga}},
  \bibinfo{journal}{Rev. Mod. Phys.} \textbf{\bibinfo{volume}{82}},
  \bibinfo{pages}{1225} (\bibinfo{year}{2010}),
  \urlprefix\url{http://link.aps.org/doi/10.1103/RevModPhys.82.1225}.

\bibitem[{\citenamefont{Petrov et~al.}(2005)\citenamefont{Petrov, Salomon, and
  Shlyapnikov}}]{Petrov_Salomon_05}
\bibinfo{author}{\bibfnamefont{D.~S.} \bibnamefont{Petrov}},
  \bibinfo{author}{\bibfnamefont{C.}~\bibnamefont{Salomon}}, \bibnamefont{and}
  \bibinfo{author}{\bibfnamefont{G.~V.} \bibnamefont{Shlyapnikov}},
  \bibinfo{journal}{J.~Phys.~B} \textbf{\bibinfo{volume}{38}},
  \bibinfo{pages}{S645} (\bibinfo{year}{2005}),
  \urlprefix\url{http://stacks.iop.org/0953-4075/38/i=9/a=014}.

\bibitem[{\citenamefont{Feshbach}(1962)}]{Feshbach_62}
\bibinfo{author}{\bibfnamefont{H.}~\bibnamefont{Feshbach}},
  \bibinfo{journal}{Ann.~Phys.} \textbf{\bibinfo{volume}{19}},
  \bibinfo{pages}{287 } (\bibinfo{year}{1962}), ISSN \bibinfo{issn}{0003-4916},
  \urlprefix\url{http://www.sciencedirect.com/science/article/pii/000349166290221X}.

\bibitem[{\citenamefont{D\"{u}rr et~al.}(2006)\citenamefont{D\"{u}rr, Volz,
  Syassen, Bauer, Hansis, and Rempe}}]{Derr_Volz_06}
\bibinfo{author}{\bibfnamefont{S.}~\bibnamefont{D\"{u}rr}},
  \bibinfo{author}{\bibfnamefont{T.}~\bibnamefont{Volz}},
  \bibinfo{author}{\bibfnamefont{N.}~\bibnamefont{Syassen}},
  \bibinfo{author}{\bibfnamefont{D.~M.} \bibnamefont{Bauer}},
  \bibinfo{author}{\bibfnamefont{E.}~\bibnamefont{Hansis}}, \bibnamefont{and}
  \bibinfo{author}{\bibfnamefont{G.}~\bibnamefont{Rempe}},
  \bibinfo{journal}{AIP Conf. Proc.} \textbf{\bibinfo{volume}{869}},
  \bibinfo{pages}{278} (\bibinfo{year}{2006}),
  \urlprefix\url{http://link.aip.org/link/?APC/869/278/1}.

\bibitem[{\citenamefont{Landau and Lifshitz}(1977)}]{Landau_Lifshitz_77}
\bibinfo{author}{\bibfnamefont{L.~D.} \bibnamefont{Landau}} \bibnamefont{and}
  \bibinfo{author}{\bibfnamefont{E.~M.} \bibnamefont{Lifshitz}},
  \emph{\bibinfo{title}{Quantum Mechanics (Non-relativistic Theory)}},
  vol.~\bibinfo{volume}{3} (\bibinfo{publisher}{Pergamon Press},
  \bibinfo{address}{Tarrytown, New York}, \bibinfo{year}{1977}).

\bibitem[{\citenamefont{Claussen et~al.}(2003)\citenamefont{Claussen,
  Kokkelmans, Thompson, Donley, Hodby, and Wieman}}]{Claussen_Kokkelmans_03}
\bibinfo{author}{\bibfnamefont{N.~R.} \bibnamefont{Claussen}},
  \bibinfo{author}{\bibfnamefont{S.~J. J. M.~F.} \bibnamefont{Kokkelmans}},
  \bibinfo{author}{\bibfnamefont{S.~T.} \bibnamefont{Thompson}},
  \bibinfo{author}{\bibfnamefont{E.~A.} \bibnamefont{Donley}},
  \bibinfo{author}{\bibfnamefont{E.}~\bibnamefont{Hodby}}, \bibnamefont{and}
  \bibinfo{author}{\bibfnamefont{C.~E.} \bibnamefont{Wieman}},
  \bibinfo{journal}{Phys. Rev. A} \textbf{\bibinfo{volume}{67}},
  \bibinfo{pages}{060701} (\bibinfo{year}{2003}),
  \urlprefix\url{http://link.aps.org/doi/10.1103/PhysRevA.67.060701}.

\bibitem[{\citenamefont{Kohn}(1959)}]{Kohn_59}
\bibinfo{author}{\bibfnamefont{W.}~\bibnamefont{Kohn}}, \bibinfo{journal}{Phys.
  Rev.} \textbf{\bibinfo{volume}{115}}, \bibinfo{pages}{809}
  (\bibinfo{year}{1959}).

\bibitem[{\citenamefont{Auerbach}(1994)}]{Auerbach_94}
\bibinfo{author}{\bibfnamefont{A.}~\bibnamefont{Auerbach}},
  \emph{\bibinfo{title}{Interacting Electrons and Quantum Magnetism}}
  (\bibinfo{publisher}{Springer}, \bibinfo{address}{Berlin},
  \bibinfo{year}{1994}).

\bibitem[{\citenamefont{Jaksch et~al.}(1998)\citenamefont{Jaksch, Bruder,
  Cirac, Gardiner, and Zoller}}]{jaksch_bruder_98}
\bibinfo{author}{\bibfnamefont{D.}~\bibnamefont{Jaksch}},
  \bibinfo{author}{\bibfnamefont{C.}~\bibnamefont{Bruder}},
  \bibinfo{author}{\bibfnamefont{J.~I.} \bibnamefont{Cirac}},
  \bibinfo{author}{\bibfnamefont{C.~W.} \bibnamefont{Gardiner}},
  \bibnamefont{and} \bibinfo{author}{\bibfnamefont{P.}~\bibnamefont{Zoller}},
  \bibinfo{journal}{Phys. Rev. Lett.} \textbf{\bibinfo{volume}{81}},
  \bibinfo{pages}{3108} (\bibinfo{year}{1998}).

\bibitem[{\citenamefont{Lercher et~al.}(2011)\citenamefont{Lercher, Takekoshi,
  Debatin, Schuster, Rameshan, Ferlaino, Grimm, and N\"agerl}}]{lercherAD2011}
\bibinfo{author}{\bibfnamefont{A.~D.} \bibnamefont{Lercher}},
  \bibinfo{author}{\bibfnamefont{T.}~\bibnamefont{Takekoshi}},
  \bibinfo{author}{\bibfnamefont{M.}~\bibnamefont{Debatin}},
  \bibinfo{author}{\bibfnamefont{B.}~\bibnamefont{Schuster}},
  \bibinfo{author}{\bibfnamefont{R.}~\bibnamefont{Rameshan}},
  \bibinfo{author}{\bibfnamefont{F.}~\bibnamefont{Ferlaino}},
  \bibinfo{author}{\bibfnamefont{R.}~\bibnamefont{Grimm}}, \bibnamefont{and}
  \bibinfo{author}{\bibfnamefont{H.-C.} \bibnamefont{N\"agerl}},
  \bibinfo{journal}{Eur. Phys. J. D} pp. \bibinfo{pages}{1--7}
  (\bibinfo{year}{2011}).

\bibitem[{\citenamefont{Genz and Malik}(1980)}]{Genz_Malik_80}
\bibinfo{author}{\bibfnamefont{A.}~\bibnamefont{Genz}} \bibnamefont{and}
  \bibinfo{author}{\bibfnamefont{A.}~\bibnamefont{Malik}},
  \bibinfo{journal}{J.~Comput.~Appl.~Math.} \textbf{\bibinfo{volume}{6}},
  \bibinfo{pages}{295 } (\bibinfo{year}{1980}), ISSN \bibinfo{issn}{0377-0427},
  \urlprefix\url{http://www.sciencedirect.com/science/article/pii/0771050X8090039X}.

\bibitem[{\citenamefont{Stroud}(1971)}]{Stroud_71}
\bibinfo{author}{\bibfnamefont{A.}~\bibnamefont{Stroud}},
  \emph{\bibinfo{title}{Approximate calculation of multiple integrals}}
  (\bibinfo{publisher}{Prentice-Hall}, \bibinfo{address}{Englewood Cliffs,
  N.~J.}, \bibinfo{year}{1971}).

\bibitem[{\citenamefont{Press et~al.}(1993)\citenamefont{Press, Teukolsky,
  Vetterling, and Flannery}}]{press1993}
\bibinfo{author}{\bibfnamefont{W.~H.} \bibnamefont{Press}},
  \bibinfo{author}{\bibfnamefont{S.~A.} \bibnamefont{Teukolsky}},
  \bibinfo{author}{\bibfnamefont{W.~T.} \bibnamefont{Vetterling}},
  \bibnamefont{and} \bibinfo{author}{\bibfnamefont{B.~P.}
  \bibnamefont{Flannery}}, \emph{\bibinfo{title}{Numerical Recipes in C: The
  Art of Scientific Computing}} (\bibinfo{publisher}{Cambridge Univ. Press},
  \bibinfo{address}{Cambridge, U.K.}, \bibinfo{year}{1993}).

\bibitem[{\citenamefont{Berntsen et~al.}(1991)\citenamefont{Berntsen, Espelid,
  and Genz}}]{Berntsen_Espelid_91}
\bibinfo{author}{\bibfnamefont{J.}~\bibnamefont{Berntsen}},
  \bibinfo{author}{\bibfnamefont{T.~O.} \bibnamefont{Espelid}},
  \bibnamefont{and} \bibinfo{author}{\bibfnamefont{A.}~\bibnamefont{Genz}},
  \emph{\bibinfo{title}{An adaptive algorithm for the approximate calculation
  of multiple integrals}} (\bibinfo{year}{1991}).

\bibitem[{\citenamefont{Genz}(1991)}]{Genz91anadaptive}
\bibinfo{author}{\bibfnamefont{A.}~\bibnamefont{Genz}}, in
  \emph{\bibinfo{booktitle}{Computing in the 90s, Proceedings of the First
  Great Lakes Computer Science Conference, Lecture Notes in Computer Science
  Volume 507}} (\bibinfo{publisher}{Springer}, \bibinfo{year}{1991}), pp.
  \bibinfo{pages}{279--292}.

\bibitem[{\citenamefont{Golub and {Van Loan}}(1996)}]{golubGH1996}
\bibinfo{author}{\bibfnamefont{G.~H.} \bibnamefont{Golub}} \bibnamefont{and}
  \bibinfo{author}{\bibfnamefont{C.~F.} \bibnamefont{{Van Loan}}},
  \emph{\bibinfo{title}{Matrix Computations}}, Johns Hopkins Studies in
  Mathematical Sciences (\bibinfo{publisher}{The Johns Hopkins University
  Press}, \bibinfo{address}{Baltimore}, \bibinfo{year}{1996}),
  \bibinfo{edition}{3rd} ed.

\bibitem[{\citenamefont{Lancaster}(1966)}]{lancaster_66}
\bibinfo{author}{\bibfnamefont{P.}~\bibnamefont{Lancaster}},
  \emph{\bibinfo{title}{Lambda-matrices and Vibrating Systems}}
  (\bibinfo{publisher}{Pergamon Press}, \bibinfo{address}{Oxford},
  \bibinfo{year}{1966}).

\end{thebibliography}

\end{document}